\documentclass[prd,aps,floatfix,nofootinbib,preprint,tightenlines,showpacs]{revtex4}
\usepackage[pdftex]{color}
\usepackage[sort&compress]{natbib}
\usepackage[colorlinks=true,linkcolor=blue,filecolor=blue,urlcolor=blue,citecolor=blue,pdftex=true,plainpages=false]{hyperref}
\usepackage{amsmath,amssymb,amsbsy,amsfonts,bm}
\usepackage{graphicx}
\usepackage[mathscr]{euscript}
\usepackage{epsfig}
\usepackage{bm}
\usepackage{psfrag}
\usepackage{rotating}
\usepackage{slashed}
\usepackage{float}
\usepackage{setspace}

                                {\end{figure}}
                                {\end{table}}

\def\rcite#1{Ref.~\cite{#1}}

\def\chpt{\raise0.4ex\hbox{$\chi$}PT}
\def\schpt{S\raise0.4ex\hbox{$\chi$}PT}
\def\rschpt{rS\raise0.4ex\hbox{$\chi$}PT}
\def\rhmschpt{rHMS\raise0.4ex\hbox{$\chi$}PT}
\def\secref#1{Sec.~\ref{sec:#1}}

\def\Secref#1{Section~\ref{sec:#1}}
\def\tabref#1{Table~\ref{tab:#1}}

\def\eqn#1{\label{eq:#1}}
\def\Equation#1{Equation~(\ref{eq:#1})}

\def\eq#1{Eq.~(\ref{eq:#1})}
\def\eqsthru#1#2{Eqs.~(\ref{eq:#1}) through (\ref{eq:#2})}
\def\eqs#1#2{Eqs.~(\ref{eq:#1}) and (\ref{eq:#2})}
\def\eqsthree#1#2#3{Eqs.~(\ref{eq:#1}), (\ref{eq:#2}) and (\ref{eq:#3})}
\def\eqsfour#1#2#3#4{Eqs.~(\ref{eq:#1}), (\ref{eq:#2}), (\ref{eq:#3}) and (\ref{eq:#4})}

\def\figref#1{Fig.~\ref{fig:#1}}
\def\figrefeq#1{Fig.\ref{fig:#1}}

\def\figrefs#1#2{Figs.~\ref{fig:#1} and \ref{fig:#2}}

\def\cD{{\cal D}}

\def\cH{{\cal H}}
\def\cI{{\cal I}}

\def\cL{{\cal L}}
\def\cM{{\cal M}}

\def\cO{{\cal O}}

\def\cQ{{\cal Q}}

\def\cS{{\cal S}}
\def\cT{{\cal T}}

\def\cV{{\cal V}}
\def\cW{{\cal W}}

\def\bar{\overline}
\def\bea{\begin{eqnarray}}
\def\eea{\end{eqnarray}}
\def\spose#1{\hbox to 0pt{#1\hss}}
\def\vslash{{v \!\!\!/}}

\def\MeV{{\rm Me\!V}}

\def\gtwid{{\,\raise.35ex\hbox{$>$\kern-.75em\lower1ex\hbox{$\sim$}}\,}}
\def\ltwid{{\,\raise.35ex\hbox{$<$\kern-.75em\lower1ex\hbox{$\sim$}}\,}}
\def\leftvec{{\raise1.5ex\hbox{$\leftarrow$}\kern-1.00em}}
\def\rightvec{{\raise1.5ex\hbox{$\rightarrow$}\kern-1.00em}}
\newcommand*{\Tr}{\ensuremath{\operatorname{Tr}}}
\newcommand*{\tr}{\ensuremath{\operatorname{tr}}}

\newcommand{\negcdot}{\negmedspace\cdot\negmedspace}

\newcommand{\vk}{\ensuremath{v\negcdot k}}
\newcommand{\Delstar}{\ensuremath{\Delta^{\raise0.18ex\hbox{${\scriptstyle *}$}}}}

\def\ie{{\it i.e.},\ }

\def\ala{{\it \`a la}}

\begin{document}

\title{\bf\large Neutral $B$ Mixing in Staggered Chiral Perturbation Theory}

\author{C. Bernard}
\affiliation{Department of Physics, Washington University, St. Louis, MO 63130, USA}

\author{(The MILC Collaboration)}
\date{\today\\ \vspace{1.2cm}}

\begin{abstract}
I calculate, at
one loop in staggered chiral perturbation theory, the
matrix elements of the complete set
of five local operators that may contribute to $B$ mixing both in
the Standard Model and in beyond-the-Standard-Model theories.
 Lattice computations of these matrix elements by the
Fermilab Lattice/MILC collaborations (and earlier by
the HPQCD collaboration) convert a light staggered quark into
a naive quark, and construct the relevant 4-quark operators
as local products of two local bilinears, each involving the naive
light quark and the heavy quark.  This particular representation
of the operators turns out to be important in the chiral calculation,
and it results in the presence of ``wrong-spin'' operators, whose contributions
however vanish in the continuum limit.   If the matrix elements
of all five operators are computed on the lattice, then 
no additional low energy constants are required to describe wrong-spin 
chiral effects.
\end{abstract}

\maketitle

\section{Introduction}

The mixing of neutral $B$ mesons provides a fertile area for precision
tests of the Standard Model.  The fact that the mixing is a second order weak process 
and is also suppressed by small CKM angles in the Standard Model makes it
sensitive to new physics.  In order to take full advantage of experimental measurements
of the mixing, one needs to determine the hadronic matrix elements of the
effective weak operators. For $B$ mixing the relevant operators are local four-quark
operators with $\Delta b=2$, where $b$ is $b$-quark number, and the relevant states
are $B_d^0$ and $\bar B_d^0$ mesons or $B_s^0$ and $\bar B_s^0$ mesons. 
A first-principle evaluation of such operator matrix elements is possible with lattice QCD.

Lattice computations usually involve an extrapolation in light
quark masses to the physical up and down masses, and always require an extrapolation in
lattice spacing $a$ to $a=0$, the continuum.  These extrapolations can be controlled by
using a version of chiral perturbation theory that includes the effects of the
discretization errors associated with the choice of lattice action.
In two  recent lattice calculations of
$B$ mixing \cite{Gamiz:2009ku,Bazavov:2012zs}, 
staggered light quarks are combined with non-staggered heavy quarks using
NRQCD \cite{NRQCD} or the Fermilab action \cite{El-Khadra:1996mp}, respectively. 
In such cases, the appropriate chiral theory
is ``rooted, heavy-meson staggered chiral perturbation theory'' (\rhmschpt) \cite{Aubin:2005aq}.

In this paper, I calculate $B$ mixing to one-loop order in \rhmschpt. Roughly speaking, I
work to leading order in the heavy-quark expansion, although I do include the large
$1/m_B$ effects: the hyperfine splitting of $B$ and $B^*$ and the flavor splitting
of $B_s$ and $B_d$.  This is a systematic approximation in the power
counting introduced by Boyd and Grinstein \cite{Boyd:1994pa} and discussed recently 
in Ref.~\cite{Bazavov:2011aa} for the
lattice calculation of heavy-light meson decay constants.
If instead one prefers a power counting strictly in $1/m_B$, which sets the splittings
to zero, it is easy to take that limit of the results given in this paper.

In the \rhmschpt\ calculation,
it is important to take into account the exact form of the lattice operator used to
approximate the continuum one.  References~\cite{Gamiz:2009ku,Bazavov:2012zs} construct
the four-quark operators as the local product of two local bilinears,%
\footnote{That is, all four fields are located at the same lattice point.}
each
formed from a heavy antiquark field and a light quark field with the ``naive'' lattice
discretization.
As proposed in Ref.~\cite{Wingate:2002fh} and discussed below,
the naive field is constructed in turn from the simulated staggered fields  (or more precisely, 
the naive propagator is constructed from the staggered propagator). Both the use of naive
fields and the local nature of the four-quark operator influence the form of the corrections
at one loop.  

It is not hard to understand the qualitative effect of the lattice locality of the four-quark
operator.  Because of lattice doubling symmetry, a (single-component) staggered quark
field actually corresponds to the 16 continuum degrees of freedom of four ``tastes''
of four-component Dirac particles.  On the lattice, the spin and taste degrees 
of freedom 
can be made explicit in position space by combining the 16 staggered components associated
with an elementary hypercube \cite{TASTE-REPRESENTATION}. For our four-quark operators,
the two light staggered quarks are tied to the same space-time point, so that their spin 
and tastes are coupled. The coupling produces undesired contributions to the operator,
with ``wrong spin'' and ``wrong taste.'' These undesired contributions appear at $\cO(1)$
in the lattice spacing.  Fortunately, in the matrix elements considered here, continuum
$SU(4)$ taste symmetry suppresses wrong-taste contributions and therefore wrong-spin
contributions.  On the lattice, $SU(4)$ taste symmetry is violated at $\cO(a^2)$, so
the undesired contributions come in at that order.  Since  $\cO(a^2)$  corrections appear
at one loop in \rhmschpt, that is the order at which we find wrong-spin, wrong-taste contributions
to $B$ mixing.

Similarly, it is clear that the effect of using naive quarks in the operators
and interpolating fields must be to sum over tastes, since  the
naive quarks have no explicit taste index. However, the details are non-trivial. It turns
out that the heavy-light meson propagator is simply an average over the initial and final
tastes, which are equal to each other. The three-point function
involves a complicated sum over tastes of the staggered quarks in the interpolating
fields and four-quark operator, and there is coupling between the spin matrices in
the operator and the taste sum.  These details play a key part in the discussion below.

A calculation in \rhmschpt\ can be thought of as ``staggering'' the corresponding continuum 
calculation, which here would be in heavy-meson chiral perturbation theory.  In 
fact, when the continuum
calculation includes partial quenching effects, it is often possible to deduce the proper
staggered version without having to recalculate explicitly
any of the diagrams (see, for example, Ref.~\cite{Aubin:2007mc}). In the current case, a
partially quenched continuum calculation does exist \cite{Detmold:2006gh}.
However, the complications due to the naive-to-staggered translation and the wrong spin-taste
contributions make it necessary to perform  
the staggered calculation from scratch.  Nevertheless,
Ref.~\cite{Detmold:2006gh} is extremely useful here, and provides a check of the current
results in the $a\to0$ limit.

The $B$ mixing matrix elements for any four-quark operator that can appear in
the Standard Model and in possible
extensions such as supersymmetry can be written in terms of the matrix elements
of the following five operators  \cite{Gabbiani:1996hi}
\begin{eqnarray}
\cO_1 &=& (\bar b \gamma^\nu L q)\; [\bar b \gamma^\nu L q]\nonumber \\
\cO_2 &=& (\bar b L q) \;[\bar b L q]\nonumber \\
\cO_3 &=& (\bar b  L q]\; [\bar b L q) \nonumber\\
\cO_4 &=& (\bar b L q)\; [\bar b  R q] \nonumber\\
\cO_5 &=& (\bar b L q]\; [\bar b R q) \ ,
\eqn{SUSY-basis}
\end{eqnarray}
where pairs of round or square parenthesis indicate how the color indices are to be contracted,
and $R$ and $L$ are the right and left projectors: $R=(1+\gamma_5)/2$ and $L=(1-\gamma_5)/2$.
Operators, $\cO_1$, $\cO_2$ and $\cO_3$ appear in the Standard Model, with $\cO_1$
(which mixes with $\cO_2$ under renormalization) governing the mass differences of the
neutral $B$ eigenstates, $\Delta M_d$ and $\Delta M_s$. Operators $\cO_4$ and $\cO_5$ appear in
extensions of the Standard Model.  Additional operators with $R\leftrightarrow L$ can
also contribute beyond the Standard Model, but parity implies
that their mixing matrix elements in QCD are equal
to those of the above operators.  In addition to parity, Fierz transformations
are needed in order to write the mixing matrix elements of any four-quark operator with these quantum numbers
in terms of those in \eq{SUSY-basis}; for a detailed explanation see Ref.~\cite{Bouchard:2011yia}.

I note that the corresponding projectors $R,L$ in
Ref.~\cite{Detmold:2006gh} do not have the factor of $1/2$, so the operators there are
differently normalized.  Since in any case unknown low energy constants will enter in
the chiral theory, this normalization difference is unimportant here.

The fact that \eq{SUSY-basis} is a complete set of operators for $B$ mixing implies that
wrong-spin contributions to the operators do not in fact lead to any new low energy
constants in the chiral theory.  Wrong-spin contributions to operator
$\cO_i$ merely lead to the appearance of the low energy constants associated with 
operators $\cO_{j\not=i}$ in the one-loop expression for the $\cO_i$ matrix element.
Thus, a staggered lattice calculation that computes the matrix elements of all the operators
in \eq{SUSY-basis} will not suffer from increased systematic or statistical errors due to
the wrong-spin issue.  Existing calculations \cite{Gamiz:2009ku,Bazavov:2012zs} study
the matrix element $\cO_1$ exclusively.  In the case of Ref.~\cite{Gamiz:2009ku}, the
one-loop contributions of wrong-spin operators were not known at the time, 
so one presumably should include some
additional systematic error in their result. In the case of Ref.~\cite{Bazavov:2012zs},
it was not possible to make a complete study of this effect because the matrix elements
of the other operators were not computed.  
However, an associated systematic error was estimated.  

In the relevant staggered simulations, the fourth-root of the fermion determinant is
taken in order to eliminate the four-fold multiplicity of tastes in the sea.
The rooted theory then suffers from nonlocal violations of unitarity 
at non-zero lattice spacing~\cite{Prelovsek05,BGS06}. However, there
are strong theoretical arguments~\cite{Bernard06,Shamir04,Shamir06,BGS08},
as well as other analytical and numerical
evidence~\cite{SharpePoS06,KronfeldPoS07,GoltermanPoS08,Bazavov:2009bb,Donaldetal11},
that the local, unitary theory of QCD is recovered in the continuum
limit.  Furthermore, it is straightforward to take rooting into account in the
chiral theory.  One simply needs to multiply each sea quark  loop by a factor of $1/4$ 
\cite{Aubin:2003mg,Aubin:2003uc}.  This can be done  either
by following the quark flow \cite{QUARK-FLOW} to locate the loops, 
or --- more systematically --- by replicating the sea quarks $n_r$
times and taking $n_r=1/4$ in the result of the chiral calculation \cite{Bernard06,BGS08}.
Since I will need to work out the quark flows in any case, I use the former method below.

The remainder of this paper is organized as follows.  \Secref{basics} briefly reviews the basics
of \rhmschpt, focusing in particular on those aspects that will be important here.
In \secref{naive-staggered}, I discuss the connection between naive and 
staggered quarks, and how
it influences the structure of the four-quark operators and the interpolating heavy-light meson 
fields. The calculation of the one-loop diagrams is detailed in \secref{NLO}.
I also briefly explain why taste-violations coming from mixing under renormalization do
not need to be considered at this order.
\Secref{results} compiles 
the final formulae  
for the chiral and continuum extrapolation of the matrix elements  
of the operators defined in \eq{SUSY-basis}; corresponding
results for the B (``bag'') parameters are collected in Appendix~\ref{sec:Bparams}.  
I conclude in \secref{conclusions}
and make some additional comments about existing and future lattice computations
of $B$ mixing. A preliminary account of the current calculation appears in 
Ref.~\cite{Bernard:2012rn}.

Though I denote heavy quarks as $b$ quarks and heavy mesons
as $B$ mesons throughout,  the current calculation in \rhmschpt\ also applies to the
local matrix elements in neutral
$D$ mixing, with the usual caveat that the omitted $1/m_Q$ terms ($Q$ is a generic heavy quark) 
are larger in that case.  However, long distance contributions are 
presumably much more important in the $D$ case \cite{Wolfenstein:1985ft} than in the $B$ case.
Such contributions
are beyond the scope of this work, and are likely to be difficult to compute on the lattice.
See however Ref.~\cite{Christ:2010gi}
 for a lattice approach to long-distance effects in the kaon system.

\section{Basics of  rHMS\raise0.4ex\hbox{$\chi$}PT}
\label{sec:basics}

Here, I give a brief summary of
some of the basic features and definitions from staggered chiral perturbation
theory, both of heavy-light mesons and of light mesons (``pions'').
In this summary, I follow Ref.~\cite{Aubin:2007mc} fairly closely, but
adapt the notation slightly, to make it more similar to that of Ref.~\cite{Detmold:2006gh}.
 The reader is
referred to the literature 
\cite{Lee:1999zxa,Aubin:2003mg,Aubin:2003uc,Sharpe:2004is,Aubin:2005aq,Aubin:2007mc} 
for more details. For convenience in making connection to Ref.~\cite{Detmold:2006gh},
I write the Lagrangian and do the perturbative calculations in Minkowski space.

Let $P^{(b)}_q$ be the field that
annihilates the pseudoscalar meson containing a heavy quark $b$ and a light quark $\bar q$ (the $\bar
B_0$ for $q=d$), while $ P^{\ast (b)}_{\mu,q}$ does the same for the 
vector meson ($\bar B^\ast_0$ for $q=d$). 
To take advantage of heavy-quark spin symmetry, pseudoscalar and vector fields
are combined in the field
\begin{equation}
  H^{(b)}_q = \frac{1 + \vslash}{2}\left[ \gamma^\mu_{(M)} P^{\ast (b)}_{\mu,q}
    + i \gamma_5^{(M)} P^{(b)}_q\right]\ ,\eqn{hdef}
\end{equation}
which destroys a meson, while
\begin{equation}
  \overline{H}^{(b)}_q = 
\left[ \gamma^\mu_{(M)} P^{\ast (b)\dagger}_{\mu,q}
    + i \gamma_5^{(M)} P^{(b)\dagger}_q\right]\frac{1 + \vslash}{2}
\ ,\eqn{hbardef}
\end{equation}
creates a meson.
Here $v$ is the meson velocity, and the  
$(M)$ on $\gamma^\mu_{(M)}$ and
$\gamma_5^{(M)}$
indicates that they are 
Minkowski-space  matrices: $\gamma^0_{(M)}= \gamma^0$, $\gamma^j_{(M)}= i\gamma^j$,
 and $\gamma_5^{(M)}= \gamma_5$, with
$\gamma^\mu$ and $\gamma_5$ the Euclidean (Hermitian) Dirac matrices. 
The label $q$ indicates the ``flavor-taste''
index of the light quark in the meson. 
For $n$ flavors of light quarks, $q$ can take on $4n$ values. 
Later, I will write $q$ as separate flavor ($x$) and taste ($a$) indices, 
$q\to (x,a)$.

Under $SU(2)$ heavy-quark spin symmetry, 
the heavy-light field
transforms as
\begin{eqnarray}
  H^{(b)} &\to & S H^{(b)}\ , \nonumber\\
  \overline{H}^{(b)} &\to & \overline{H}^{(b)}S^{\dagger}\ ,
\end{eqnarray}
with $S\in SU(2)$, 
while under
the $SU(4n)_L\times SU(4n)_R$ chiral symmetry,
\begin{eqnarray}
  H^{(b)} &\to & H^{(b)} \mathbb{U}^{\dagger}\ ,\nonumber\\
  \overline{H}^{(b)} &\to & \mathbb{U}\overline{H}^{(b)}\ ,
\end{eqnarray}
with $\mathbb{U}\in SU(4n)$ defined below. 
We keep the light flavor and taste indices implicit here. 

The light mesons are combined in a Hermitian field $\Phi(x)$.
For $n$ staggered flavors, $\Phi$ is a $4n \times 4n$
matrix given by:
\begin{eqnarray}\label{eq:Phi}
  \Phi = \left( \begin{array}{cccc}
      U  & \pi^+ & K^+ & \cdots \\*
      \pi^- & D & K^0  & \cdots \\*
      K^-  & \bar{K^0}  & S  & \cdots \\*
      \vdots & \vdots & \vdots & \ddots \end{array} \right)\ .
\end{eqnarray}
I show the $n=3$ portion of $\Phi$ explicitly, and in fact detailed
final results below will assume $n=3$. 
Each entry in \eq{Phi} is a  $4\!\times\!4$ matrix, written
in terms of the 16 Hermitian basis elements of the Clifford taste algebra. 
It is convenient to take
the generators of this algebra to be 
$\xi_\mu=\gamma_\mu^*$ \cite{TASTE-REPRESENTATION}, where $*$ denotes complex conjugation.
Thus we write, 
for example,
\begin{eqnarray}
U &=& \sum_{\Xi=1}^{16} U_\Xi \Gamma^*_\Xi\ , \eqn{Ufielddef0}\\
  \Gamma_\Xi &=& \{ \gamma_5,\;
i\gamma_{\mu}\gamma_5,\;
  \sigma_{\mu\nu}\, (\mu<\nu),\; \gamma_{\mu},\;
  I\}\ ,\eqn{Gamma_Xi}
\end{eqnarray}
with 
$\sigma_{\mu\nu}\equiv (i/2)[\gamma_{\mu},\gamma_{\nu}]$.

It is useful to divide the indices $\Xi$ into
pairs of indices: $\Xi\to (\rho,t_\rho)$, where $\rho$ labels the 
SO(4) representation (P,A,T,V,I) and $t_\rho$ labels the element within each
representation.  Thus $t_\rho$ runs from 1 to $N_\rho$, where $N_\rho$ is the
dimension of each representation (1,4,6,4,1, respectively).
We then can write
\begin{equation}
U = \sum_{\rho}\sum_{t_\rho=1}^{N_\rho} U_{\rho,t_\rho}\; \Gamma^*_{\rho,t_\rho}. \eqn{Ufielddef}
\end{equation}

The component fields of the flavor-neutral
elements  of $\Phi$ (namely $U_{\rho,t_\rho}$, $D_{\rho,t_\rho}$, \dots) are real; 
the other (flavor-charged)
fields ($\pi^+_{\rho,t_\rho}$,  $K^0_{\rho,t_\rho}$, \dots) are complex. 

 The
mass matrix is the $4n\times 4n$ matrix
\begin{eqnarray}
  \cM = \left( \begin{array}{cccc}
      m_u I  & 0 &0  & \cdots \\*
      0  & m_d I & 0  & \cdots \\*
      0  & 0  & m_s I  & \cdots\\*
      \vdots & \vdots & \vdots & \ddots \end{array} \right),
\end{eqnarray}
where 
the portion shown is again for the $n=3$ case.

  From $\Phi$ one constructs the 
unitary chiral field $\Sigma = \exp [i\Phi/f]$,
with $f$ the tree-level pion decay constant. In our normalization, $f \sim f_\pi \cong 131\ \MeV$.
Terms involving the heavy-lights are conveniently written using 
$\sigma \equiv \sqrt{\Sigma} = \exp[ i\Phi / 2f ]$. 
These fields transform trivially under
the $SU(2)$ spin symmetry, 
while under $SU(4n)_L\times SU(4n)_R$ we have
\begin{eqnarray}
  \Sigma \to  L\Sigma R^{\dagger}\,,\qquad&&\qquad
  \Sigma^\dagger \to  R\Sigma^\dagger L^{\dagger}\,,\\*
  \sigma \to  L\sigma \mathbb{U}^{\dagger} = \mathbb{U} \sigma R^{\dagger}\,, \qquad&&\qquad
  \sigma^\dagger \to R \sigma^\dagger \mathbb{U}^{\dagger} = \mathbb{U} \sigma^\dagger L^{\dagger}\,, 
  \label{eq:Udef}
\end{eqnarray}
with global transformations $L\in SU(4n)_L$ and $R\in SU(4n)_R$.
The transformation $\mathbb{U}$, defined by \eq{Udef}, is
a function of $\Phi$ and therefore 
of the coordinates.

It is convenient to define objects involving the $\sigma$ field that transform 
only with $\mathbb{U}$ and $\mathbb{U}^\dagger$. 
The two 
possibilities with a single derivative are
\begin{eqnarray}
  \mathbb{V}_{\mu} & = & \frac{i}{2} \left[ \sigma^{\dagger} \partial_\mu
   \sigma + \sigma \partial_\mu \sigma^{\dagger}   \right] \ ,
 \\
  \mathbb{A}_{\mu} & = & \frac{i}{2} \left[ \sigma^{\dagger} \partial_\mu
   \sigma - \sigma \partial_\mu \sigma^{\dagger}   \right] \ .
\end{eqnarray}
$\mathbb{V}_{\mu}$ transforms like a vector field under the $SU(4n)_L\times SU(4n)_R$ 
chiral symmetry and, 
when combined with the derivative, 
can form a 
covariant derivative acting on the heavy-light field or its conjugate: 
\begin{eqnarray}\label{eq:Ddef}
	(H^{(b)} \leftvec D_\mu)_q  = H^{(b)}_{q'} \leftvec D^{{q'}q}_\mu  
	&\equiv& \partial_\mu H^{(b)}_q + i H^{(b)}_{q'}\mathbb{V}_{\mu}^{{q'}q}\ , 
	\nonumber \\
	(\rightvec D_\mu \overline{H}^{(b)})_q  = 
	\rightvec D^{q{q'}}_\mu \overline{H}^{(b)}_{q'} 
	&\equiv& \partial_\mu \overline{H}^{(b)}_q - 
	i \mathbb{V}_{\mu}^{q{q'}} \overline {H}^{(b)}_{q'}\ ,
\end{eqnarray}
with implicit sums over repeated indices.
The covariant derivatives and $\mathbb{A}_\mu$
transform under the chiral symmetry as
\begin{eqnarray}\label{eq:Dtransf}
	H^{(b)} \leftvec D_\mu &\to&  (H^{(b)} \leftvec D_\mu )\mathbb{U}^\dagger\ , \nonumber \\
	\rightvec D_\mu \overline{H}^{(b)} &\to&  \mathbb{U} (\rightvec D_\mu \overline{H}^{(b)})\ ,\nonumber \\
	 \mathbb{A}_\mu &\to&  \mathbb{U} \mathbb{A}_\mu \mathbb{U}^\dagger\ .
\end{eqnarray}

We can write the leading order (LO) chiral Lagrangian as 
\begin{equation}\label{eq:Lcont}
  \cL_{LO} = \cL_{\rm pion}+ \cL_{\rm HL} \ ,
\end{equation}
where $\cL_{\rm pion}$ is the standard staggered chiral perturbation theory
(\schpt) Lagrangian for the light-light mesons, 
and $\cL_{\rm HL}$
is the contribution of the heavy-lights. 
In Minkowski space, 
we have
\begin{eqnarray}
	\cL_{\rm pion} & = & \frac{f^2}{8} \Tr(\partial_{\mu}\Sigma 
  \partial^{\mu}\Sigma^{\dagger}) + 
  \frac{1}{4}\mu f^2 \Tr(\cM\Sigma+\cM\Sigma^{\dagger})
  \nonumber\\&&{}
  - \frac{2m_0^2}{3}(U_I + D_I + S_I+\ldots)^2 - a^2 \cV \ ,
  \label{eq:Lpion}\\
  -\cV & = & C_1
  \Tr(\xi_5\Sigma\xi_5\Sigma^{\dagger})
  +C_3\frac{1}{2} \sum_{\nu}[ \Tr(\xi_{\nu}\Sigma
    \xi_{\nu}\Sigma) + h.c.] \nonumber \\*&&
  {}+C_4\frac{1}{2} \sum_{\nu}[ \Tr(i\xi_\nu\xi_5\Sigma
    i\xi_\nu\xi_5\Sigma) + h.c.]
   +C_6\ \sum_{\mu<\nu} \Tr(\xi_{\mu\nu}\Sigma
  \xi_{\mu\nu}\Sigma^{\dagger})  \nonumber \\*&&
   {}+C_{2V}\frac{1}{4} \sum_{\nu}[ \Tr(\xi_{\nu}\Sigma)
   \Tr(\xi_{\nu}\Sigma)
    + h.c.]
   +C_{2A}\frac{1}{4} \sum_{\nu}[ \Tr(i\xi_\nu\xi_5 \Sigma)
   \Tr(i\xi_\nu\xi_5\Sigma)
    + h.c.] \nonumber \\*
  && {}+C_{5V}\frac{1}{2} \sum_{\nu} \Tr(\xi_{\nu}\Sigma)
  \Tr(\xi_{\nu}\Sigma^{\dagger})
  + C_{5A}\frac{1}{2}\sum_{\nu}\Tr(i\xi_\nu\xi_5\Sigma)
  \Tr(i\xi_\nu\xi_5\Sigma^{\dagger}) \label{eq:VSigma} \ ,\\
  \cL_{\rm HL}  &=&  -i \Tr(\overline{H}^{(b)} H^{(b)}\; v\negcdot \leftvec D )
  + g_\pi \Tr(\overline{H}^{(b)}H^{(b)}\gamma^{\mu}_{(M)}\gamma_5^{(M)} 
  \mathbb{A}_{\mu}) \ .
\label{eq:L-HL}
\end{eqnarray}
Here $\Tr$ denotes a trace over flavor-taste indices and, 
where relevant, 
Dirac indices.  The product  $\overline{H}^{(b)}H^{(b)}$ is treated as a matrix in flavor-taste space: 
$(\overline{H}^{(b)}H^{(b)})_{qq'} \equiv \overline{H}^{(b)}_qH^{(b)}_{q'}$.
The covariant derivative $\leftvec D$ acts only on the field 
immediately preceding it.
For convenience, I work with diagonal fields ($U$, $D$, \dots) and leave
the anomaly ($m_0^2$) term explicit in \eq{Lpion}. 
We can take 
$m^2_0\to\infty$ and go to the physical basis ($\pi^0$, $\eta$, \dots)
at the end of the calculation
\cite{SHARPE-SHORESH}.

At tree level, the light-light meson composed of quarks of flavor $x$ and $y$,
and with $SO(4)$ taste representation $\rho$, is
\begin{equation}
       M_{xy,\rho}^2  = \mu (m_x + m_y) + a^2\Delta_\rho\;. \eqn{mpi2}
\end{equation}
Here $\Delta_\rho$ is the taste splitting,
which can be expressed in terms of $C_1$,
$C_3$,
$C_4$ and $C_6$
in \eq{VSigma} \cite{Aubin:2003mg}.
The residual $SO(4)$ taste
symmetry \cite{Lee:1999zxa} at this order
implies that the mesons within a given taste representation 
are degenerate in mass.

I now list some key expressions from the Feynman
rules given in Ref.~\cite{Aubin:2007mc}, but adapted to the current notation. 
Using separate indices for flavor ($x,y$) and taste ($a,a',c,c'$),
the (quark-line) connected pion propagator in Minkowski space is
\begin{equation}\label{eq:PropConn}
        \Bigl\{\Phi^{xy}_{aa'}\Phi^{yx}_{c'c}\Bigr\}_{\rm conn}(p) 
 = \sum_{\rho}
        \frac{i }
{p^2 - M_{xy,\rho}^2 + i\epsilon}
\bigg[\sum_{t_\rho}  \Gamma^{\rho,t_\rho}_{a'a}\;  \Gamma^{\rho,t_\rho}_{cc'}\bigg] \ .
\end{equation}
Here, I have used the fact that $\Gamma^{\rho,t_\rho}$ is Hermitian
is to replace the complex conjugation in \eq{Ufielddef} by interchange of indices on the
right-hand side, which will be convenient later.

Similarly, the disconnected (hairpin) propagator is
\begin{equation}\label{eq:PropDisc}
        \Bigl\{\Phi^{xx}_{aa'}\Phi^{yy}_{c'c}\Bigr\}_{\rm disc}(p) \equiv 
\sum_\rho  \cD^\rho_{xx,yy}\bigg[\sum_{t_\rho}
\Gamma^{\rho,t_\rho}_{a'a}\;  \Gamma^{\rho,t_\rho}_{cc'} \bigg] \ , 
\end{equation}
where
 \begin{equation}\label{eq:Disc}
\cD^\rho_{xx,yy} = -i\delta'_\rho \frac{i}{(p^2-M_{X,\rho}^2+i\epsilon)}
 \frac{i}{(p^2-M_{Y,\rho}^2+i\epsilon)}
  \frac{(p^2-M_{U,\rho}^2)(p^2-M_{S,\rho}^2)}
       {(p^2-M_{\eta,\rho}^2+i\epsilon)(p^2-M_{\eta',\rho}^2+i\epsilon)}\ ,
\end{equation}
with the hairpin strength $\delta'_\rho$ given by
\begin{equation}\label{eq:dp-def}
  \delta_\rho' = \begin{cases}
    a^2 \delta'_V, &\rho=V\ \textrm{(taste\ vector);}\\*
    a^2 \delta'_A, &\rho=A\ \textrm{(taste\ axial-vector);}\\*
    4m_0^2/3, &\rho=I\ \textrm{(taste\ singlet);}\\*
    0, &\rho=T,P\ \textrm{(taste\ 
    tensor or pseudoscalar)}\ .
  \end{cases}
\end{equation}
$X$ and $Y$ denote valence mesons made from $x\bar x$ or $y\bar y$ quarks, respectively,
with $M_X$ and $M_Y$ their masses. For
the sea mesons, the masses $M_U$ and $M_S$ do not include
the mixing effects of the hairpins. The re-diagonalized
states after including the hairpins
are $\eta$ and $\eta'$.
For concreteness I have assumed the 2+1 case: $m_u=m_d$.  

The propagators for the heavy-light mesons are
\begin{eqnarray}
        \Bigl\{P^{(b)}_{xa} P^{(b)\dagger}_{yc}\Bigr\}(k) &=& \frac{i\delta_{ac}\delta_{xy}}
        {2(\vk + i\epsilon)}\ , \label{eq:Bbarprop}\\
        \Bigl\{P^{*(b)}_{\mu xa} P^{*(b)\dagger}_{\nu yc}\Bigr\}(k) &=& 
        \frac{-i\delta_{ac}\delta_{xy}(g_{\mu\nu} - v_\mu v_\nu)}
        {2(\vk -\Delstar + i\epsilon)} \ \label{eq:Bbarstarprop},
\end{eqnarray}
where $\Delstar$ is the $B^*$-$B$ mass splitting.
The $\bar B\,\bar B^*\,\pi$ vertex (including the $i$ from $\exp(i\cL)$) is:
\begin{equation}\label{eq:Bbar-Bbarstar-pi}
\frac{g_\pi}{f}\,\left( P^{(b)\dagger}_{xa} \, P^{*(b)}_{\mu yc} - 
P^{*(b)\dagger}_{\mu xa}\, P^{(b)}_{yc} \, 
\right) \, \partial^\mu \Phi_{ca}^{yx} \ ,
\end{equation}
where the repeated indices $a,\ x,\ y,\ c,$ and $\mu$ are summed.

For $B$ mixing, we also need corresponding fields that destroy and create mesons
with $\bar b$ quarks, \ie $B_0$-like and
$B^\ast_0$-like mesons. These fields and their interactions
can be obtained from the previous ones
using charge conjugation \cite{Grinstein:1992qt}.
The individual meson
fields are indicated by, for example, 
$P^{\ast (\bar b)}_{\mu,q}$ and
$P^{(\bar b)}_q$.  (Note that the light quark label $q$ does not distinguish between
quarks and antiquarks.) The combined fields are
\begin{eqnarray}
  H^{(\bar b)}_q &=& 
\left[ \gamma^\mu_{(M)} P^{\ast (\bar b)}_{\mu,q}
    + i \gamma_5^{(M)} P^{(\bar b)}_q\right]\frac{1 - \vslash}{2}\ ,\eqn{hdef-bbar} \\
  \overline{H}^{(\bar b)}_q &=& \frac{1 - \vslash}{2}
\left[ \gamma^\mu_{(M)} P^{\ast (\bar b)\dagger}_{\mu,q}
    + i \gamma_5^{(M)} P^{(\bar b)\dagger}_q\right]
\ .\eqn{hbardef-bbar}
\end{eqnarray}
The propagators for the $P^{\ast (\bar b)}_{\mu,q}$ and
$P^{(\bar b)}_q$ fields are the same as those for the $P^{\ast (b)}_{\mu,q}$ and
$P^{(b)}_q$ fields, \eqs{Bbarprop}{Bbarstarprop}.
The $BB^*\pi$ vertex is
\begin{equation}\label{eq:B-Bstar-pi}
\frac{g_\pi}{f}\,\left( P^{(\bar b)}_{xa} \, P^{*(\bar b)\dagger}_{\mu yc} - 
P^{*(\bar b)}_{\mu xa}\, P^{(\bar b)\dagger}_{yc} \, 
\right) \, \partial^\mu \Phi_{ca}^{yx} \ .
\end{equation}

\section{Translating from Naive to Staggered Quarks}
\label{sec:naive-staggered}

The naive light quark action may be rewritten as four copies of the staggered
action:
\begin{equation}
\eqn{omega}
\Psi(x) = \Omega(x)\;\underline{\chi}(x)\;; \qquad \Omega(x) = 
\gamma_0^{x_0}\,\gamma_1^{x_1}\,\gamma_2^{x_2}\,\gamma_3^{x_3},
\end{equation}
where $\Psi(x)$ is the naive quark field and $\underline{\chi}(x)$ is a ``copied'' staggered
field, with each Dirac component $\underline{\chi}{}_i $ separately having the staggered
action.  I call ``copy symmetry'' the
$SU(4)$ that acts on the copy index $i$.
Unlike the $SU(4)$ vector taste symmetry, which acts on individual 
staggered fields (written in the spin-taste basis) and is exact for an interacting theory
only in the continuum limit,
copy symmetry is an exact lattice symmetry.  Thus the propagator of
a copied staggered field is
\begin{equation}
\eqn{copied-prop}
\langle \underline{\chi}{}_i(x)\, \underline{\bar\chi}{}_{i'}(y)\rangle = \delta_{i,i'}\;
\langle\chi(x)\, \bar\chi(y)\rangle,
\end{equation}
where $\chi$ is the normal (uncopied) staggered field.
This implies
\begin{equation}
\eqn{naive-prop}
\langle \Psi(x)\, \bar\Psi(y)\rangle =\Omega(x)\,\Omega^\dagger(y)
\langle\chi(x)\, \bar\chi(y)\rangle \ .
\end{equation}
In the simulations using staggered quarks, the naive field is never constructed
{\it per se}\/; instead \eq{naive-prop} is used to translate staggered
propagators  into naive propagators \cite{Wingate:2002fh}.

An interpolating field $\cH(x)$ for a $B_q$ meson is
\begin{equation}
\eqn{interp-field}
\cH(x) = \bar b(x)\,\gamma_5\, \Psi(x) = \bar b(x)\,\gamma_5\, \Omega(x)\,\underline{\chi}(x)\;.
\end{equation}
I assume that in practical applications $\cH(x)$ will always
be summed over a time-slice, either explicitly, or implicitly by using translation
invariance.

To leading order in $a$, $b(x)$ varies smoothly (up to gauge transformation)
between neighboring spatial sites, but $\underline{\chi}$ does not, due to taste doubling.
On the other hand, in the spin-taste basis, which we arrive at by summing the staggered
fields over hypercubes, the staggered fields are smooth on the doubled lattice.  We
are thus led to focus on the average of $\cH(x)$ over a spatial cube. Let 
 $x=(t,{\bm x})$ with ${\bm x}=2{\bm y}$ even, and let
$\eta=(\eta_0,{\bm \eta})$ be a 4-vector with
all components 0 or 1. For $t$ even ($t=2\tau$) the averaged field is
\begin{eqnarray}
\cH^{\rm(av)}(t,{\bm x}) &=&  \frac{1}{8} \sum_{\bm \eta}\bar b(t,{\bm x}+{\bm \eta})\,\gamma_5\, 
   \Omega(2\tau,{\bm \eta})\,\underline{\chi}(2\tau,2{\bm y}+{\bm \eta}) \nonumber \\
&\cong&\frac{1}{8}\; \bar b(t,{\bm x})\,\gamma_5  \sum_{\bm \eta} 
            \Omega({\bm \eta})\,\underline{\chi}(2\tau,2{\bm y}+{\bm \eta}) \nonumber \\
&\cong& \frac{1}{16}\; \bar b(t,{\bm x})\,\gamma_5\; \sum_{\eta}
\big[\Omega(\eta)\,\underline{\chi}(2\tau+\eta_0,2{\bm y}+{\bm \eta})+\nonumber \\
	&&\hspace{28mm}+(-1)^{\eta_0}\,\Omega(\eta)\,
\underline{\chi}(2\tau+\eta_0,2{\bm y}+{\bm \eta}) \big]\ .
\eqn{H-av0} 
\end{eqnarray}
Inserted gauge links for gauge invariance of point-split quantities are implicit.
For $t$ odd ($t=2\tau+1$), the result is the same except the term
on the last line of \eq{H-av0} changes sign.  Using the fact that
$(-1)^{\eta_0}\,\Omega(\eta) = \gamma_5\gamma_0 \,\Omega(\eta)\, \gamma_0\gamma_5$,
it is not hard to see that this second term just gives the usual staggered
oscillating (in time)
state with opposite parity.  I have dropped higher order terms in $a$ coming from
the variation of the heavy-quark field over the cube.
For simplicity,
we simply assume from now on that all components of
$x$ are even ($x_\mu=2y_\mu$), and that the
oscillating state has been removed by the fitting procedure.
We then have
\begin{equation}
\cH^{\rm(av)}(x) \to  \frac{1}{16}\; \bar b(x)\,\gamma_5\, \sum_{\eta}
\Omega(\eta)\,\underline{\chi}(2y+\eta) \eqn{H-av1} \ .
\end{equation}

We now convert to a spin-taste basis for the staggered fields.  The standard
construction for a single staggered field is \cite{TASTE-REPRESENTATION}
\begin{equation}
q^{\alpha a}(y) =  \frac{1}{8}\; \sum_{\eta}
\Omega^{\alpha a}(\eta)\,\chi(2y+\eta)\ , \eqn{Kluberg} 
\end{equation}
where $\alpha$ is a spin index and $a$ is a taste index. As is well known 
\cite{Golterman:1984cy,Sharpe:1993ng,Lepage:2011vr},
this decomposition is correct only to lowest order in $a$ and
generates a spurious ${\cal O}(a)$ term in the spin-taste action,
but it is good enough for our purposes.
Here we need a copied
version:
\begin{equation}
q^{\alpha a}_i(y) =  \frac{1}{8}\; \sum_{\eta}
\Omega^{\alpha a}(\eta)\,\underline{\chi}{}_i(2y+\eta)\ . \eqn{CopyKluberg} 
\end{equation}
With spin indices implicit, \eq{H-av1} then becomes
\begin{equation}
\cH^{\rm(av)}(x) \to  \frac{1}{2}\; \bar b(x)\,\gamma_5\, q^a_i(y)\, \delta^a_i\ ,
\eqn{H-av2}
\end{equation}
where repeated indices are summed. With \eq{copied-prop}, this implies that the
contraction of $\cH$ with $\cH^\dagger$ ({\it i.e.}, the heavy-light propagator) is automatically
averaged  over tastes:
\begin{equation}
\langle \cH(x)\,\cH^\dagger(x')\rangle \sim  \frac{1}{4}\; \langle\bar b(x)\gamma_5 q^a(y)\; 
\bar q^a(y')\gamma_5 b(x')\rangle\ ,
\eqn{H-prop}
\end{equation}
where a sum over $a$ is implicit.

Analysis of the four-quark operator is more complicated because the two bilinears from which it is constructed
are not separately summed over space; only the four-quark operator is summed.  
However, we can write it in
terms of separately summed bilinears by using the identities
\begin{eqnarray}
\frac{1}{256}\sum_K \tr\!\left(\Omega(\eta)\,K\, \Omega^\dagger(\eta)\, K\right)\, 
\tr\!\left(\Omega(\eta')\,K\, \Omega^\dagger(\eta')\, K\right)
&=&   \delta_{\eta,\eta'}\ ,\eqn{eta-ident} \\
\frac{1}{4}\tr\!\left(\Omega(\eta)\,K\, \Omega^\dagger(\eta)\, K\right)\,\Omega(\eta) = K\Omega(\eta)K\ .
\end{eqnarray}
Here $K$  is any of the 16 independent Hermitian matrices $\Gamma_\Xi$ in \eq{Gamma_Xi}, which obey $K^2=I$.
We get, for operator $\cO_n = \bar b\Gamma_n\Psi\;\bar b\Gamma'_n\Psi$:
\begin{eqnarray}
\cO^{(av)}_n(x) &=& \frac{1}{8}\sum_{\bm\eta} \bar b(t,{\bm 2y+\eta}) \Gamma_n \Psi(t,{\bm 2y+\eta}) 
\; \bar b(t,{\bm 2y+\eta}) \Gamma'_n \Psi(t,{\bm 2y+\eta}) \nonumber \\
&\to& \frac{1}{4} \sum_K (\bar b  \Gamma_n K q^c_k \;
\bar b  \Gamma'_n K q^d_\ell)\; K_{ck} K_{d\ell}\ .
\eqn{On-av}
\end{eqnarray}
Here we have dropped the ``wrong parity'' part, which does not contribute if oscillating
terms are removed by the fitting procedure.
Note that contributions with $K\not=I$ have incorrect spin ($\Gamma_nK\otimes\Gamma'_nK$ instead of
$\Gamma_n\otimes\Gamma'_n$), and coupling of taste ($c,d$) and copy
($k,\ell$) indices. 

There are $\cO(a)$ and higher corrections to \eqs{H-av2}{On-av}, 
coming from the
variations of the heavy quark field over the spatial cube and from corrections
to the spin-taste construction in position space. As 
discussed in \secref{taste-breaking}, however, such terms
do not contribute to non-analytic terms
in chiral perturbation theory until NNLO.

Using \eqs{H-av2}{On-av}, copy symmetry (\eq{copied-prop}) implies
\begin{equation}
\langle \cH^{(av)\dagger}\; \cO^{(av)}_n\; \cH^{(av)\dagger} \rangle \propto \langle D^{ca}\, D^{de}  \rangle\;  K_{ca} K_{de}\ ,
\eqn{On-props}
\end{equation}
where $D^{ca}$ is the quark propagator (in a given background) for taste $a$ into taste $c$. 
If taste symmetry is exact, $\langle D^{ca} D^{de}  \rangle\propto \delta_{ac} \delta_{ed}$, and
only the correct spin ($K=I$) contributes.  Thus
the desired matrix element will be obtained in the continuum limit.

 At one loop, however, taste violations allow $\langle D^{ac} D^{ed}  \rangle$ not to
be proportional to $\delta_{ac} \delta_{ed}$, 
and terms with incorrect
spins can contribute.
For example, the taste-violating hairpin with vector taste can give a term proportional to $\xi^\mu_{ca\phantom{e}}\! \xi^\mu_{de}= \gamma^\mu_{ac\phantom{e}}\! \gamma^\mu_{ed}$.  Then, since $tr(\gamma^\mu K) = 4\delta_{K,\gamma_\mu}$,
the spin of the operator is $\Gamma_n \gamma_\mu \otimes
\Gamma'_n \gamma_\mu $ instead  of $\Gamma_n \otimes \Gamma'_n $

\Equation{On-av} may be simplified by taking advantage of the exact $SO(4)$ taste symmetry of the staggered chiral theory at one loop.
Within any
$SO(4)$ multiplet  $\kappa\in\{P, A, T, V, I\}$ with dimension $N_\kappa>1$  ({\it e.g.}, the 
vector-taste multiplet V with $N_V=4$), the value of all one-loop diagrams 
would be unchanged if we replaced any multiplet element in the taste factors
$K_{ck}K_{d\ell}$ with another element from the same
multiplet.%
\footnote{Although indices $k$ and $\ell$ are copy, not taste, indices at this point,
they will become taste indices \ala\ \eq{On-props} shortly.}  
We therefore write $K=\Gamma^{\kappa,t_\kappa}$,
where $t_\kappa$ labels
the element within multiplet $\kappa$.
Replacing the sum over $K$ in \eq{On-av} with a double
sum over $\kappa,t_\kappa$, and using the $SO(4)$ symmetry, we then have
\begin{eqnarray}
\cO^{(av)}_n(x) 
&\to& \frac{1}{4} \sum_{\kappa,t'_\kappa} (\bar b  \Gamma_n \Gamma^{\kappa,t'_\kappa} q^c_k \;
\bar b  \Gamma'_n \Gamma^{\kappa,t'_\kappa} q^d_\ell)\; \frac{1}{N_\kappa}\sum_{t_\kappa} 
\Gamma^{\kappa,t_\kappa}_{ck} \Gamma^{\kappa,t_\kappa}_{d\ell}\ .
\eqn{On-av-simple}
\end{eqnarray}
Within a given multiplet, this decouples the sum over spins from the sum over tastes.

Thus, for a given continuum operator $\cO_n$, the plan is to 
calculate the one-loop diagrams for each of the operators
\begin{equation}
\cO_n^{\kappa} \equiv \sum_{t'_\kappa}\bar b  \Gamma_n \Gamma^{\kappa,t'_\kappa} q^c_k \;
\bar b  \Gamma'_n \Gamma^{\kappa,t'_\kappa} q^d_\ell  
\eqn{On-root}
\end{equation}
between external (interpolating) fields 
$(\cH^{\dagger})^{a}_i$ and  $(\cH^{\dagger})^{e}_j$, where, from \eq{H-av2},
\begin{equation}
(\cH^{\dagger})^{a}_i\equiv \bar q^a_i\gamma_5 b \ .
\eqn{H-root}
\end{equation}
Each diagram for operator $\cO_n^{\kappa}$ then gets an additional factor $\tilde F_\kappa$
coming from \eqs{On-av-simple}{H-av2}, where
\begin{equation}
\tilde F_\kappa \equiv \frac{1}{16N_\kappa}\sum_{t'_\kappa}\Gamma^{\kappa,t'_\kappa}_{ck}
\Gamma^{\kappa,t'_\kappa}_{d\ell}\delta_{ai}\delta_{ej}\;.
\eqn{factor1}
\end{equation}
Whether explicitly indicated or not, all repeated indices  will then be summed; 
this includes taste and copy indices ($a,c,d,e$ and
$i,j,k,\ell$) as well as the indices with dual, spin-taste meaning 
($\kappa,t_\kappa,t'_\kappa$).

\section{Calculation of one-loop diagrams for $O_n$}
\label{sec:NLO}

\subsection{Procedure}
\label{sec:Procedure}

We now set up one-loop \rhmschpt\ for the operators described above.
We follow  Ref.~\cite{Detmold:2006gh}
as much as possible,
but must take into account the complications of copy and taste indices.  
It is convenient first to express the operators $\cO_n^{\kappa}$, given
in \eq{On-root} in
terms of the basis of \eq{SUSY-basis}.
From the relations among operators listed for example in \rcite{Bouchard:2011yia} we find
\begin{eqnarray}
\cO_1^{P}&=& \cO_1 \ ,\nonumber \\
\cO_1^{A}&=& -8\cO_2 -8\cO_3\ ,\nonumber \\
\cO_1^{T}&=& -6\cO_1  \ ,\eqn{O1-kappa}\\
\cO_1^{V}&=& 8\cO_2 +8\cO_3\ ,\nonumber \\
\cO_1^{I}&=& \cO_1 \ ,\nonumber
\end{eqnarray}
\begin{eqnarray}
\cO_2^{P}&=& \cO_2 \ ,\nonumber \\
\cO_2^{A}&=& -\cO_1 \ ,\nonumber \\
\cO_2^{T}&=& -2\cO_2-4\cO_3  \ ,\eqn{O2-kappa}\\
\cO_2^{V}&=& \cO_1 \ ,\nonumber \\
\cO_2^{I}&=& \cO_2 \ ,\nonumber
\end{eqnarray}
\begin{eqnarray}
\cO_3^{P}&=& \cO_3 \ ,\nonumber \\
\cO_3^{A}&=& -\cO_1 \ ,\nonumber \\
\cO_3^{T}&=& -4\cO_2-2\cO_3  \ ,\eqn{O3-kappa}\\
\cO_3^{V}&=& \cO_1 \ ,\nonumber \\
\cO_3^{I}&=& \cO_3 \ ,\nonumber
\end{eqnarray}
\begin{eqnarray}
\cO_4^{P}&=& -\cO_4 \ ,\nonumber \\
\cO_4^{A}&=& -2\cO_5\ , \phantom{-2\cO_x^2}\nonumber \\
\cO_4^{T}&=& 0 \ ,\eqn{O4-kappa}\\
\cO_4^{V}&=& -2\cO_5 \ ,\nonumber \\
\cO_4^{I}&=& \cO_4 \ ,\nonumber
\end{eqnarray}
\begin{eqnarray}
\cO_5^{P}&=& -\cO_5 \ ,\nonumber \\
\cO_5^{A}&=& -2\cO_4 \ ,\phantom{-2\cO_x^2}\nonumber \\
\cO_5^{T}&=& 0 \ , \eqn{O5-kappa}\\
\cO_5^{V}&=& -2\cO_4 \ ,\nonumber \\
\cO_5^{I}&=& \cO_5 \ .\nonumber
\end{eqnarray}

The chiral representatives of the standard operators on the right hand side of 
\eqsthru{O1-kappa}{O5-kappa}
are given
in \rcite{Detmold:2006gh}. There, the only relevant quantum number of the light quarks is
their flavor, and both bilinears have the same flavor, which is
labeled $q$.  Here we also need to
label the taste and, for the moment, the copy index of the light quarks, and these are not in general the same for
both light quarks in the operator.  
So we adopt the notation $q\to x,c,k$, where $x$ labels the quark flavor only, 
$c$ (or other letters near
the beginning of the alphabet) labels the quark taste, and $k$ (or other letters near
the middle of the alphabet) labels the quark copy.
From
\cite{Detmold:2006gh}, we then have
\begin{eqnarray}
 O^{xck;xd\ell}_{1} &=&  \beta_{1} \left [ 
            \Big( \sigma P^{(b)\dagger} \Big)_{x,c,k}
            \Big( \sigma P^{(\bar{b})} \Big)_{x,d,\ell}
          + \Big( \sigma P^{\ast (b)\dagger}_{\mu} \Big)_{x,c,k}
            \Big( \sigma P^{\ast (\bar{b}),\mu} \Big)_{x,d,\ell}
                 \right ] \,,
\nonumber\\
 O^{xck;xd\ell}_{2(3)} &=&  \beta_{2(3)} 
            \Big( \sigma P^{(b)\dagger} \Big)_{x,c,k}
            \Big( \sigma P^{(\bar{b})} \Big)_{x,d,\ell} 
            +  \beta_{2(3)}^\prime
            \Big( \sigma P^{\ast (b)\dagger}_{\mu} \Big)_{x,c,k}
            \Big( \sigma P^{\ast (\bar{b}),\mu} \Big)_{x,d,\ell}\,, \eqn{chiral-ops1}
\\
 O^{xck;xd\ell}_{4(5)} &=&  \frac{\beta_{4(5)}}{2}  \left [
            \Big( \sigma P^{(b)\dagger} \Big)_{x,c,k}
            \Big( \sigma^{\dagger} P^{(\bar{b})} \Big)_{x,d,\ell}  +
            \Big( \sigma^{\dagger} P^{(b)\dagger} \Big)_{x,c,k}
            \Big( \sigma P^{(\bar{b})} \Big)_{x,d,\ell} \right ]
\nonumber\\
&&           + \frac{\beta_{4(5)}^\prime}{2} \left [
            \Big( \sigma P^{\ast (b)\dagger}_{\mu} \Big)_{x,c,k}
            \Big( \sigma^{\dagger} P^{\ast (\bar{b}),\mu} \Big)_{x,d,\ell} +
            \Big( \sigma^{\dagger} P^{\ast (b)\dagger}_{\mu} \Big)_{x,c,k}
            \Big( \sigma P^{\ast (\bar{b}),\mu} \Big)_{x,d,\ell}\right ]\,.
\nonumber
\end{eqnarray}
The method used to obtain these operators in Ref.~\cite{Detmold:2006gh} is a standard
spurion analysis.  The factors of $\sigma$ and $\sigma^\dagger$ are present in order
make the light-quark spurions, which transform by 
left or right chiral rotations in \eq{SUSY-basis}, into objects that transform with 
$\mathbb{U}$ (defined in \eq{Udef}) and can combine with the heavy-meson fields
$\overline{H}^{(b)}_q$ and $H^{(\bar b)}_q $ to make invariants.  Although many insertions
of Dirac matrices are possible in forming the invariants, they all reduce down to the
simple forms in \eq{chiral-ops1} when expressed in terms of 
$P^{(b)\dagger}$, $P^{(\bar{b})}$, $P^{\ast (b)\dagger}_{\mu}$, and $P^{\ast (\bar{b})}_\mu$.
As pointed out in Ref.~\cite{Grinstein:1992qt}, this follows from heavy-quark spin 
symmetry, which relates the amplitude for $B$--$\bar B$ mixing to that of $B^\ast$--$\bar B^*$.

 In \eq{chiral-ops1}, I have used the fact that we are only interested in the
parity-even part of these operators to set the two coefficients Detmold and Lin call
$\beta_{4(5)}$ and $\hat \beta_{4(5)}$ simply to
$\beta_{4(5)}/2$. So where they have $\beta_{4(5)}+\hat \beta_{4(5)}$ we will have simply
$\beta_{4(5)}$, and similarly for $\beta^\prime_{4(5)}$ and $\hat \beta^\prime_{4(5)}$.
We use the Latin $O$ for the chiral operators to distinguish them from quark-level operators $\cO$.
The external interpolating $\bar B$ and $B$ fields are taken to be, respectively,
\begin{equation}
P^{(b)}_{x,a,i}\hspace{8mm}{\rm and}\hspace{8mm}  P^{(\bar b)\dagger}_{x,e,j}\;.
\eqn{ext-fields1}
\end{equation}

Strictly speaking, there should be a second set of terms on the right
hand sides in \eq{chiral-ops1} in which the pairs of indices $c,k$ and $d,\ell$ 
are interchanged.
These come about because the light quark field in either of the bilinears of the operator
can be the one that creates the light antiquark in the $\bar B$ or annihilates the light quark
in the $B$.  However, since \eq{factor1} is symmetric under this interchange, the extra terms
give identical results to the ones we have already, and 
therefore can be dropped at this point. Note that the $\beta_i$ are low energy constants
with arbitrary normalization.

We can now use copy symmetry to simplify the equations, and ultimately eliminate the copy
indices entirely.  However, this cannot be done without taking into account the quark flow
through the diagrams: copy symmetry works at the level of the light quark propagators,
and not on meson propagators {\it per se}.  The point is that a given light quark in
an external meson field can end up in either a $(b)$- or $(\bar b)$-labeled meson field
in \eq{chiral-ops1}, and the copy symmetry has a different effect on the diagram
in the two cases.  If the light quark in the external $P^{(b)}_{x,a,i}$ field
contracts with the quark in a  $(b)$-labeled meson  
field (which has taste and copy indices $c,k$),
then the combination of copy symmetry (which gives $\delta_{ik}$) and the $\delta_{ai}$ in \eq{factor1} forces $k=a$.
On the other hand, if the same light quark contracts with the quark in a  $(\bar b)$-labeled 
meson  field 
(indices $d,\ell$), then we have $\ell=a$.  So $\tilde F_\kappa$ will end up
with a different taste structure in the two cases.  
For convenience, I prefer to rename the taste
indices ($c\leftrightarrow d$) in the second case so that $\tilde F_\kappa$ remains
the same.  In doing so I adopt the convention that the taste-$a$ quark in the
external $(b)$-labeled  field always contracts with the taste-$c$ quark in the operator, but that
 taste-$c$ quark may be in either the  $(b)$- or the $(\bar b)$-labeled meson 
field of the operator.
Similarly, the taste-$e$ quark in the external $(\bar b)$ field always contracts
with the taste-$d$ quark in the operator.

Keeping the above points in mind, we now eliminate the copy indices completely
from the calculation.
With the exception of the  wave-function renormalization diagrams, which 
of course do not involve the
four-quark operators  at all, the final procedure is as follows:
\begin{itemize}
\item[1.\ ]{} For a particular operator $\cO_n$ of interest, we first write the related
operators $\cO_n^{\kappa}$ as linear combinations of the standard operators, following 
in \eqsthru{O1-kappa}{O5-kappa}. It will be convenient then to define $\beta_n^{(\kappa)}$ and  
$\beta_n^{\prime(\kappa)}$ as the $\beta$ and $\beta^\prime$ corresponding to operator  
$\cO_n^{\kappa}$. \tabref{beta} for $\beta_n^{(\kappa)}$ follows immediately from  
\eqsthru{O1-kappa}{O5-kappa}. 

\begin{table}[th]
\caption{Values  of $\beta_n^{(\kappa)}$. 
For $\beta_n^{\prime(\kappa)}$, simply put primes on all
entries in the table, with the understanding that
 $\beta_1^{\prime}= \beta_1$ (see \eq{chiral-ops1}).}
\label{tab:beta}
\begin{tabular}{|@{\hspace{4mm}}c@{\hspace{4mm}}|c@{\hspace{8mm}}c@{\hspace{8mm}}c@{\hspace{8mm}}c@{\hspace{8mm}}c|}
\cline{2-6}
\multicolumn{1}{c|}{}&\multicolumn{5}{c|}{$\kappa$} \\
\cline{2-6}
\multicolumn{1}{c|}{$n$}  &\rule{0pt}{2.5ex}$P$ &$A$ &$T$ &$V$ &$I$ \\
\hline
1 & $\beta_1$  & $-8\beta_2 -8\beta_3$ & $-6\beta_1$   & $8\beta_2 +8\beta_3$ & $\beta_1$  \\
2& $\beta_2$  & $-\beta_1$  & $-2\beta_2-4\beta_3$   & $\beta_1$  & $\beta_2$   \\ 
3& $\beta_3$  & $-\beta_1$  & $-4\beta_2-2\beta_3$   & $\beta_1$  & $\beta_3$  \\ 
4& $-\beta_4$  & $-2\beta_5$  & $0$  & $-2\beta_5$  & $\beta_4$  \\ 
5& $-\beta_5$  & $-2\beta_4$  & $0$   & $-2\beta_4$  & $\beta_5$  \\ 
\hline
\end{tabular}
\end{table}

\item[2.\ ]{} We then calculate the chiral diagrams, using copy-free versions
of \eq{chiral-ops1} for the chiral operators, namely
\begin{eqnarray}
 O^{xc;xd}_{1} &=&  \beta_{1} \left [ 
            \Big( \sigma P^{(b)\dagger} \Big)_{x,c}
            \Big( \sigma P^{(\bar{b})} \Big)_{x,d}
          + \Big( \sigma P^{\ast (b)\dagger}_{\mu} \Big)_{x,c}
            \Big( \sigma P^{\ast (\bar{b}),\mu} \Big)_{x,d}
                 \right ]  \hspace{3mm}[{\bf or\ } c\leftrightarrow d],
\nonumber\\
 O^{xc;xd}_{2(3)} &=&  \beta_{2(3)} 
            \Big( \sigma P^{(b)\dagger} \Big)_{x,c}
            \Big( \sigma P^{(\bar{b})} \Big)_{x,d} 
            +  \beta_{2(3)}^\prime
            \Big( \sigma P^{\ast (b)\dagger}_{\mu} \Big)_{x,c}
            \Big( \sigma P^{\ast (\bar{b}),\mu} \Big)_{x,d}
\hspace{3mm}[{\bf or\ } c\leftrightarrow d], \nonumber
\\
 O^{xc;xd}_{4(5)} &=&  \frac{\beta_{4(5)}}{2}  \left [
            \Big( \sigma P^{(b)\dagger} \Big)_{x,c}
            \Big( \sigma^{\dagger} P^{(\bar{b})} \Big)_{x,d} +
            \Big( \sigma^{\dagger} P^{(b)\dagger} \Big)_{x,c}
            \Big( \sigma P^{(\bar{b})} \Big)_{x,d} \right ]
\eqn{chiral-ops2}\\
&&\hspace{-7mm}           + \frac{\beta_{4(5)}^\prime}{2} \left [
            \Big( \sigma P^{\ast (b)\dagger}_{\mu} \Big)_{x,c}
            \Big( \sigma^{\dagger} P^{\ast (\bar{b}),\mu} \Big)_{x,d} +
            \Big( \sigma^{\dagger} P^{\ast (b)\dagger}_{\mu} \Big)_{x,c}
            \Big( \sigma P^{\ast (\bar{b}),\mu} \Big)_{x,d}\right ]\hspace{3mm}[{\bf or\ } c\leftrightarrow d].
\nonumber
\end{eqnarray}
The external interpolating fields are 
\begin{equation}
P^{(b)}_{x,a}\hspace{8mm}{\rm and}\hspace{8mm}  P^{(\bar b)\dagger}_{x,e}\;.
\eqn{ext-fields2}
\end{equation}
For a diagram with a given quark flow,
one should use either the explicit version of the operators in \eq{chiral-ops2}
or the alternative  $c\leftrightarrow d$ forms to ensure that the 
external taste-$a$ light quark contracts with the taste-$c$ light quark in
the operators (which also guarantees that tastes $e$ and $d$ contract).

\item[3.\ ]{} Each diagram is then multiplied by the overall factor
\begin{equation}
F_\kappa \equiv \frac{1}{16N_\kappa}\sum_{t_\kappa}\Gamma^{\kappa,t_\kappa}_{ca}
\Gamma^{\kappa,t_\kappa}_{de}\;,
\eqn{factor2}
\end{equation}
and the repeated taste ($a,c,d,e$) and spin ($\kappa$) indices
are summed. 
\end{itemize}

\subsection{Calculation}
\label{sec:Calculation}

We are now ready to
compute  the one-loop diagrams in \rhmschpt\ using
the Feynman rules given above in 
\eqsthru{PropConn}{Bbar-Bbarstar-pi} and \eq{B-Bstar-pi}.
Because of the complications due to
taste and copy indices, it is not possible in general simply to modify
the continuum results of
 Ref.~\cite{Detmold:2006gh} to insert staggered corrections, as in Ref.~\cite{Aubin:2007mc}. We must calculate most chiral diagrams from scratch. The exception
is the wave-function renormalization (parameterized below by the function ${\cal W}$), which is 
simple enough that the modification process (``staggering'') works.
In the wave-function case, the naive-to-staggered translation gives no complications because,
from  \eq{H-prop}, it only requires setting initial and final tastes equal and averaging
over them.  The taste-averaging has no effect because 
discrete taste symmetry (shift symmetry)
implies
that a two-point function is in any case proportional to the identity in taste space.%
\footnote{With the exact momentum space taste construction \cite{Golterman:1984cy}, this 
statement is
true to all orders in $a$, as can be seen most easily by 
using the formulation of shift symmetry 
in Ref.~\cite{BGS08}. The construction is built into staggered chiral theory, so the
statement is also true to all orders in  \rhmschpt. The fact that we have used the
(approximate) position space
taste construction \cite{TASTE-REPRESENTATION} in the translation from naive to staggered
operators is irrelevant, since for our purposes only the lowest order translation is
needed.}%

\begin{figure}[t]
\begin{center}
\begin{tabular}{c c}
\includegraphics[width=0.45\textwidth]{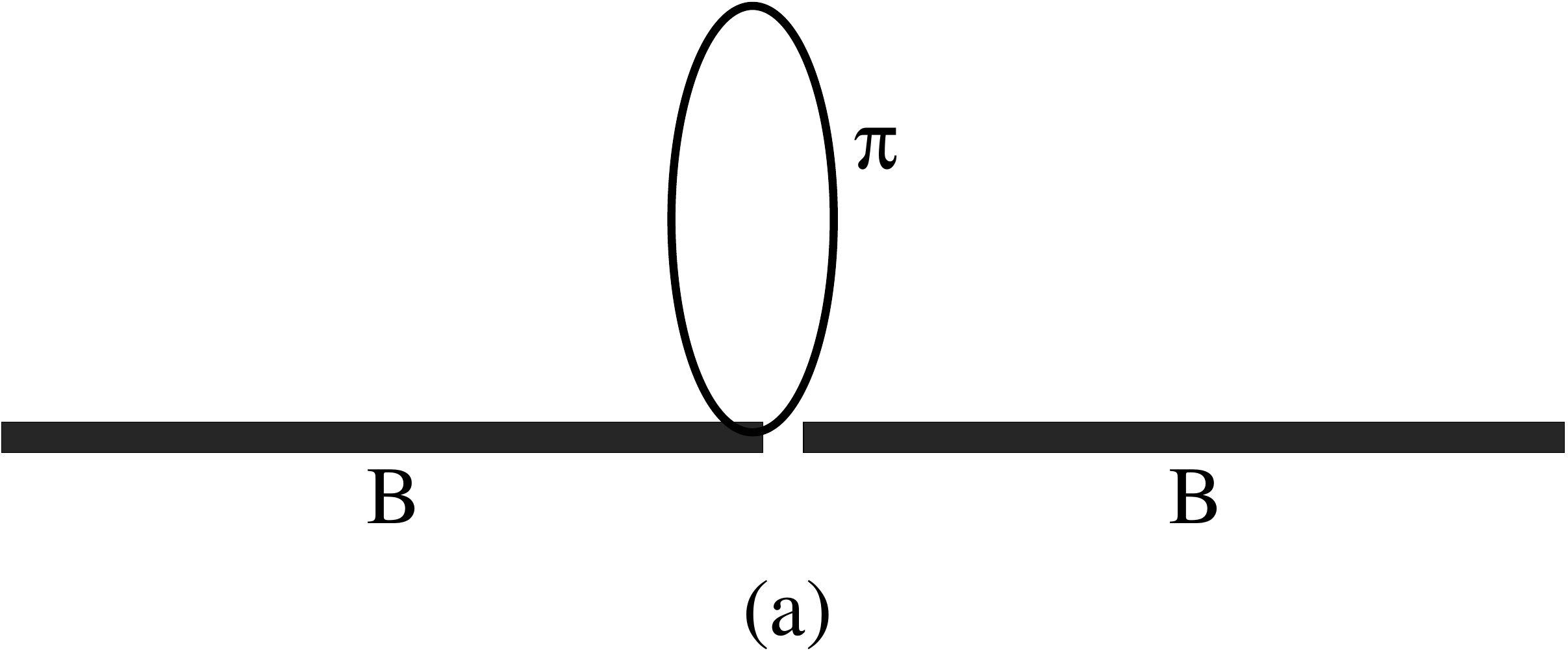}
&\hspace{7mm}
\includegraphics[width=0.45\textwidth]{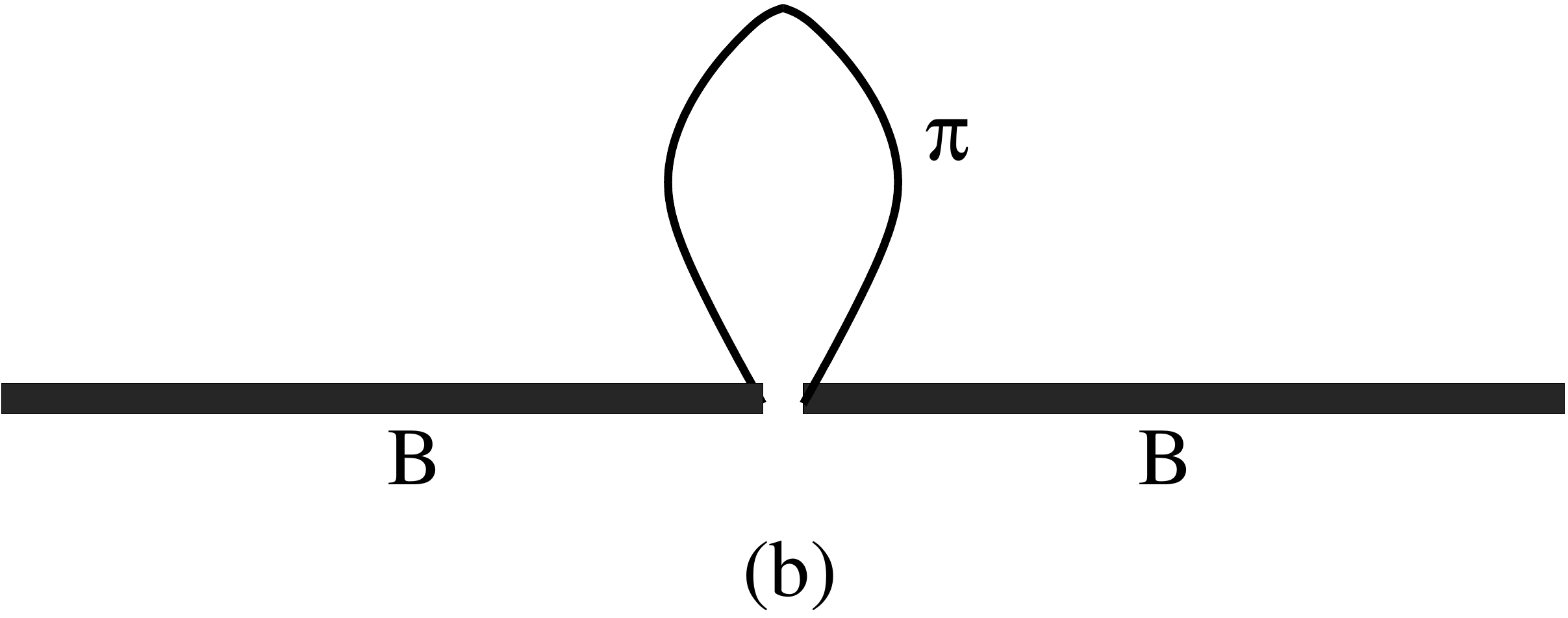}
\end{tabular}
\end{center}
\caption{Meson-level tadpole graphs for the $B$-$\bar B$ mixing matrix element. 
In (a), the
pion fields that are contracted both come from the same factor of $\sigma$ in \eq{chiral-ops2},
while in (b) they come from different $\sigma$ factors. 
For definiteness, we take the right
external line in each diagram to be the incoming $B$ meson, and the left external line, the outgoing
$\bar B$.  This means the left line is the contraction with
the $P^{(b)\dagger}$ field in \eq{chiral-ops2}, while the right line is the contraction with
the $P^{(\bar b)}$ field. 
Diagram (a) is
``factorizable'' into a product of the left- and right-hand parts (the right-hand part
is trivial here).  Diagram (b) is ``non-factorizable.'' 
There is another diagram equivalent
to (a) in which both pions come from the $\sigma$ associated with the right-hand line.
\label{fig:tad-meson}}
\end{figure}

\begin{figure}[t]
\begin{center}
\includegraphics[width=0.55\textwidth]{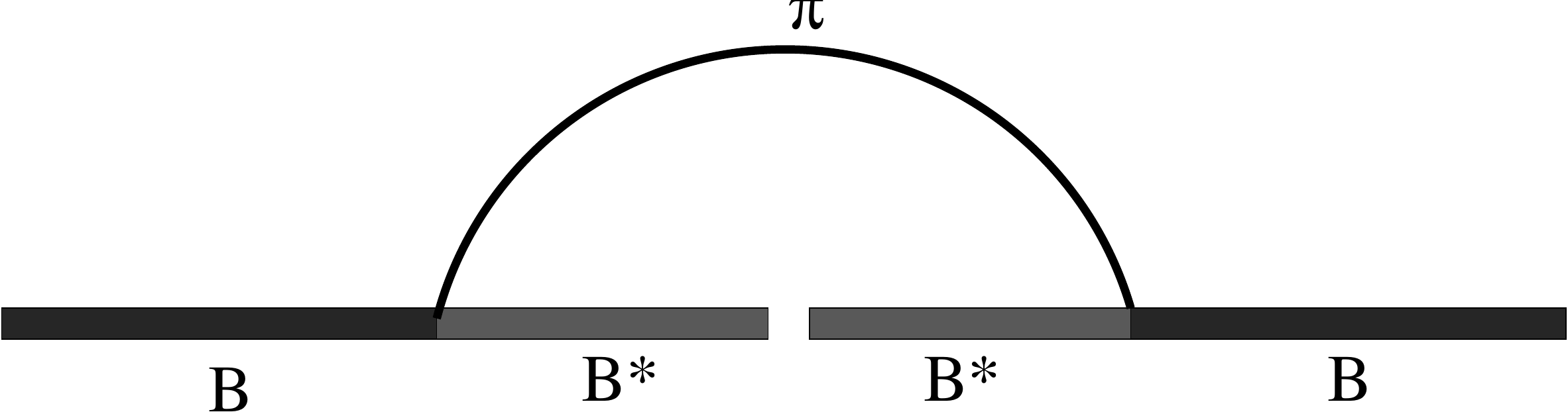}
\end{center}
\caption{The meson-level sunset graph for the $B$-$\bar B$ mixing matrix element. 
Conventions are as in \figref{tad-meson}.
\label{fig:sunset-meson}}
\end{figure}

In addition to wave-function renormalization,
there are two types of one-loop diagrams: 
tadpole graphs, \figref{tad-meson}, and
sunset diagrams, \figref{sunset-meson}. These are
parameterized by functions ${\cal T}$ and
${\cal Q}$ respectively. Contributions from incorrect spins can enter in the tadpoles
and sunset diagrams; we call such contributions $\tilde{\cal T}$,
and $\tilde{\cal Q}$. The complete matrix elements are given by
\begin{equation}
\eqn{O1tot}
\langle \overline{B}_x^0|{O}_1^x|B_x^0 \rangle =
\beta_1\left(1+
\frac{{\cal W}_{x\overline{b}}+{\cal W}_{b\overline{x}}}{2} + 
{\cal T}^{(1)}_x + \tilde{\cal T}^{(1)}_x 
+ {\cal Q}^{(1)}_x+ \tilde{\cal Q}^{(1)}_x\right)
+  \textrm{analytic terms}. 
\end{equation}
and
\begin{equation}
\langle \overline{B}_x^0|{O}_n^x|B_x^0 \rangle =
\beta_n \left(1+
\frac{{\cal
W}_{x\overline{b}}+{\cal W}_{b\overline{x}}}{2}+
{\cal T}^{(n)}_x + \tilde{\cal T}^{(n)}_x\right)
+ \beta'_n\left({\cal
Q}^{(n)}_x+ \tilde{\cal Q}^{(n)}_x\right)+ \textrm{analytic terms}, 
\eqn{O2-5tot}
\end{equation}
for $n=2,3,4,5$. Nonrelativistic normalization, which is standard in heavy-light chiral
perturbation theory, is assumed for the states $\langle \overline{B}_x^0|$
and $|B_x^0 \rangle$ in these expressions. With relativistic normalization, an extra
factor of $M_{B_x}$, the mass of the $B_x$ meson, would appear on the right-hand sides.

In the partially quenched 2+1 ($m_u=m_d\ne m_s$) case, 
staggering the result for ${\cal W}_{q\overline{b}}$ in 
Ref.~\cite{Detmold:2006gh} gives:
\begin{eqnarray} 
{\cal W}_{x\overline{b}} &=& {\cal W}_{b\overline{x}} =
\frac{ig_{B^*B\pi}^2}{f_\pi^2}\Bigg\{\frac{1}{16}\sum_{\mathscr{S},\rho}
N_\rho\,{\cal H}_{x\mathscr{S},\rho}^{\Delta^*\!+\delta_{\mathscr{S}x}}
+\frac{1}{3}\bigg[R^{[2,2]}_{X_I}\big(\{M^{(2)}_{X_I}\}
;\{\mu_I\}\big)\; \frac{\partial{{\cal
H}^{\Delta^*}_{X,I}}}{\partial m^2_{X_I}}  \nonumber
\\ && -\hspace{-3mm}\sum_{j \in
\{M^{(2)}_I\}}D^{[2,2]}_{j,X_I}\big(\{M^{(2)}_{X_I}\};\{\mu_I\}\big){\cal
H}^{\Delta^*}_{j,I} \bigg]
+a^2\delta'_{V}\bigg[R^{[3,2]}_{X_V}\big(\{M^{(3)}_{X_V}\}
;\{\mu_V\}\big)\; \frac{\partial{{\cal
H}^{\Delta^*}_{X,V}}}{\partial m^2_{X_V}}   \nonumber
\\ &&
-\hspace{-3mm}\sum_{j \in
\{M^{(3)}_V\}}D^{[3,2]}_{j,X_V}\big(\{M^{(3)}_{X_V}\};\{\mu_V\}\big){{\cal
H}^{\Delta^*}_{j,V}
 \bigg] +\big(V\rightarrow A\big)\Bigg\}}. 
\eqn{wave-function}
\end{eqnarray}
\noindent 
where the index $\rho$ runs over the taste representations (P,A,T,V,I) with degeneracies $N_\rho$,
and $\mathscr{S}$ runs over the sea mesons $u,d,s$.  
The function ${\cal H}$ is equivalent to the integral $H(m,\Delta)$ defined
in Eqs.~(A3) and (A12) of \rcite{Detmold:2006gh}. 
The subscripts on  ${\cal H}$ implicitly give the meson 
mass $m$ in $H(m,\Delta)$ by specifying
its flavor and taste.
The flavor is indicated
either by giving the flavor of the two quarks in the meson, as in $x\mathscr{S}$ in the first term
in \eq{wave-function}, or by giving the name of the meson, as in the remaining terms, where $X$
refers to the  meson made of two light valence quarks $x\bar x$ ($m_X$ is its mass).
The superscript on $\cH$ is the second argument of the function $H(m,\Delta)$.
It is the mass splitting between the
heavy-light vector meson in the chiral loop and the external
heavy-light pseudoscalar meson.  In addition to the hyperfine splitting 
$\Delta^*=M_{B^*}-M_B$, it includes a light flavor splitting 
whenever the light flavor of the vector meson in the loop
is different from the the external flavor. For the first
term in \eq{wave-function}, the vector meson has flavor $\mathscr{S}$ 
so the splitting is  $\delta_{\mathscr{S}x} \equiv M_{B_\mathscr{S}} - M_{B_x}=
2\lambda_1 \mu(m_\mathscr{S}-m_x)$, where $\lambda_1$ and $\mu$ are
 low energy constants. The constant $\lambda_1$ comes from heavy quark effective
theory, and $\mu$ relates light meson masses to quark masses, \eq{mpi2}.

For comparison, the function ${\cal H}$ is the same (up to constants) as the 
function $J$ introduced in
Ref.~\cite{Bazavov:2011aa} (Eq.~(6.17)).  Similarly the function ${\cal I}_{j,\rho}$
defined in \cite{Detmold:2006gh} and used below is the same up to constants as
the function $ \ell(m^2_{j,\rho})$ used in Refs.~\cite{Aubin:2007mc,Aubin:2003uc,Aubin:2005aq}.
The relations are:
\begin{eqnarray}
i{\cH}^\Delta_{j,\rho} &=& -\frac{3}{16\pi^2}\,J(m_{j,\rho},\Delta) \eqn{H-J} \ ,\\
i{\cI}_{j,\rho} &=& \frac{1}{16\pi^2} \ell(m^2_{j,\rho}) = \frac{1}{16\pi^2}  m^2_{j,\rho} \ln(m^2_{j,\rho}/\Lambda^2_\chi) \eqn{I-ell} \ .
\end{eqnarray}
In the limit of no splittings,
\begin{equation}
i{\cH}^0_{j,\rho} =-3i{\cI}_{j,\rho}= -\frac{3}{16\pi^2} \ell(m^2_{j,\rho})\ .
\eqn{H-J-nosplit}
\end{equation}
If one wants the $B$ mixing result in the strict $1/m_B$ power 
counting in which the splittings are set to zero,
one can simply use \eq{H-J-nosplit} for $\cH$ everywhere below.

The (Euclidean) 
residue functions $R_j^{[n,k]}$ and $D_{j,l}^{[n,k]}$ in \eq{wave-function}
are defined by \cite{Aubin:2003uc}
\begin{eqnarray} 
R^{[n,k]}_j(\{m\},\{\mu\})  &\equiv & \frac{\prod_{a=1}^k
(\mu^2_a-m^2_j)}{\prod_{i\neq j} (m^2_i-m^2_j)}\ , \nonumber \\
D^{[n,k]}_{j, l}(\{m\},\{\mu\})  &\equiv &
-\frac{d}{dm^2_l}R^{[n,k]}_j(\{m\},\{\mu\})\ . 
\end{eqnarray}
The mass combinations appearing as arguments of these functions in the 
2+1 partially quenched theory are
\begin{eqnarray} \{M_X^{(2)}\} &\equiv& \{m_\eta, m_X \}\ , \nonumber \\
    \{M_X^{(3)}\}&\equiv& \{m_\eta, m_{\eta'}, m_X \}\ , \nonumber \\
    \{\mu\}&\equiv&\{m_l,m_h\}\ . \end{eqnarray}
The tastes of these mesons ($I$, $V$, or $A$)
are indicated explicitly
in \eq{wave-function}.

The staggered heavy-light wave function renormalization is also calculated in 
Refs.~\cite{Aubin:2007mc,Aubin:2005aq} for the case where the heavy-meson splittings are neglected;
the result of adding in those splittings as explained in Ref.~\cite{Bazavov:2011aa}
agrees with
\eq{wave-function}.

The tadpole and sunset contributions are more complicated, and we must
follow the procedure outlined at the end of \secref{Procedure}.  All four taste indices
(two from the interpolating fields and two from the light quarks in the four-quark operator)
enter in a non-trivial way, and shift symmetry does not require that
all be equal.  Indeed, taste symmetry violations arising ultimately
from high-momentum gluon exchange can in general
make one pair of
tastes indices different from the other pair.  
However, if there are parts of a diagram that give simply a tree-level heavy-light
propagator, the average over taste implied by \eq{H-prop} 
can suppress the taste-changing interactions. On the other hand, in some 
diagrams for wrong-spin contributions, the overall factor, 
\eq{factor2}, 
can project onto particular taste-violating internal pion propagators. The bottom line is that
one must calculate the tadpole and sunset diagrams from first principles, and not simply
try to stagger the continuum result.%
\footnote{The rules from Ref.~\cite{Aubin:2007mc} for staggering a continuum result
would apply unchanged if, for example, we had set the two tastes
in the four-quark operator equal and averaged over them, while either fixing
the external tastes to any one value, or averaging over them. Such a definition of the
operator could be arranged if we constructed point-split bilinears with the desired tastes
within a hypercube, and then multiplied two of them appropriately. In that case, there would 
not be any wrong-spin contributions.}

\begin{figure}[t]
\begin{center}
\includegraphics[width=0.65\textwidth]{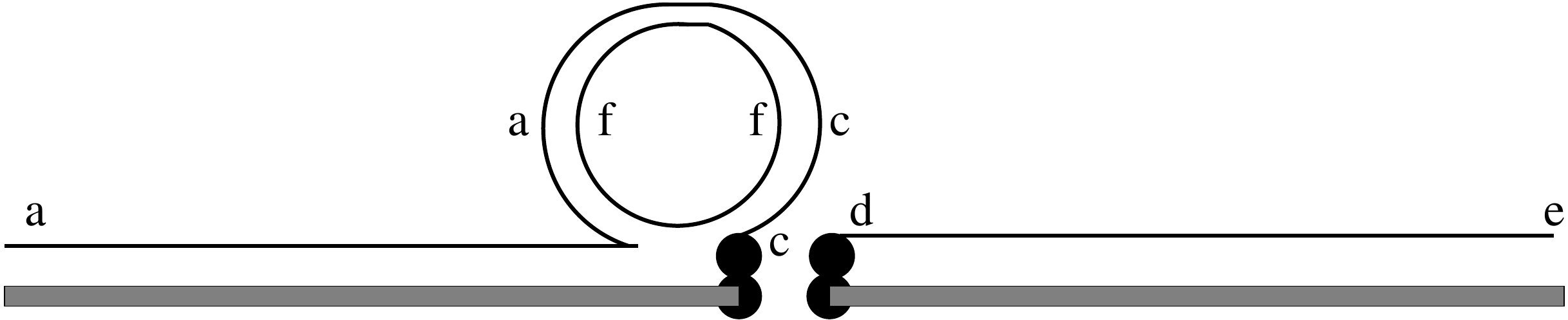}
\end{center}
\caption{Connected quark flow tadpole diagram.  This is a contribution to
\figref{tad-meson}(a). The thin lines denote staggered light quarks,
and the think, gray lines denote heavy quarks. Indices shown ($a,f,c,d,e$) are taste indices. The left pair of touching filled black circles represents the light and
heavy quark in the $P^{(b)\dagger}$; the right pair represents
the light and heavy quark in
the  $P^{(\bar b)}$. Taste conservation for the right-hand
$B$ propagator forces $d=e$. 
\label{fig:tad-quark-conn-x-s}}
\end{figure}

I start with the tadpoles.  I call a diagram ``connected'' or ``disconnected''
based on whether the internal pion propagator is connected or disconnected in the
quark flow sense of \eqs{PropConn}{PropDisc} above.
The only connected quark flow possible
for  \figref{tad-meson}(a) is shown in \figref{tad-quark-conn-x-s}.  This is a diagram with
a sea quark loop; the sea quark has taste $f$, which is summed over.  
Taste conservation for the right-hand heavy-light propagator 
(the $P^{(\bar b)}P^{(\bar b)\dagger}$ propagator) gives a factor of $\delta_{de}$.
When combined with \eq{factor2}, this implies that $\Gamma_{\kappa,t_\kappa}= I$, so this
diagram has no wrong-spin contributions.
The taste factor from \eq{PropConn} is simply $\Gamma_{af}\Gamma_{fc}= \delta_{ac}$, 
which combines
with the $\delta_{ca}$ from  \eq{factor2} (using $\Gamma_{\kappa,t_\kappa}= I$) to give
a factor of 4.  The sum over $t_\rho$ then gives $N_\rho$, the degeneracy of representation
$\rho$. Including the factor of
2 from the equivalent diagram with the loop on 
the right side of \figref{tad-quark-conn-x-s}, and
the factor of $1/4$ for the rooted sea-quark loop, we get
\begin{equation}
\cT_{x,{\rm \figrefeq{tad-quark-conn-x-s}}}^{(n)} =  -\frac{i}{16f^2}\sum_{\rho,\mathscr{S}} N_\rho\;\cI_{x\mathscr{S},\rho}\ ,
\eqn{T-conn-x-s}
\end{equation}
where $\mathscr{S}$ runs over the sea quarks $u,d,s$.  Note that
the result is independent of $n$.  Since wrong-spin contributions
to this diagram are absent, and $\beta^{(I)}_n = \beta_n$ for all $n$, 
\figref{tad-quark-conn-x-s} is simply proportional to $\beta_n$, which is 
is factored out of $\cT$ in \eqs{O1tot}{O2-5tot}. The difference in chiral structure
between the left-right operators $O_{4,5}$ and
 the left-left operators $O_{1,2,3}$ in  \eq{chiral-ops2} is not relevant
because the even terms in the expansion of $\sigma$ and $\sigma^\dagger$
are the same, and both $\pi$ fields in
\figref{tad-meson}(a) 
come from the same $\sigma$ or $\sigma^\dagger$ factor.

\begin{figure}[t]
\begin{center}
\includegraphics[width=0.65\textwidth]{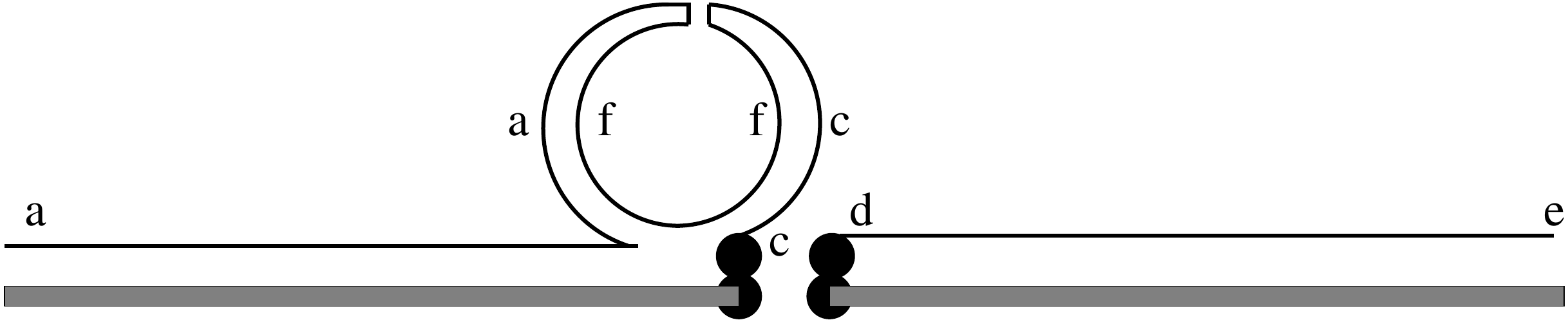}
\end{center}
\caption{Disconnected quark flow tadpole diagram.  This is a contribution to
\figref{tad-meson}(a). Taste conservation for the right-hand
$B$ propagator forces $d=e$. 
\label{fig:tad-quark-disca}}
\end{figure}

There is also a disconnected contribution to \figref{tad-meson}(a).  The quark flow diagram 
is shown in \figref{tad-quark-disca}.  Again the $\delta_{de}$ from the right-hand $B$
propagator means that only  the correct spin contributes.
The taste factor from \eq{PropDisc} is $\Gamma_{af}\Gamma_{fc}= \delta_{ac}$, which 
gives a factor 4 after using \eq{factor2} and a factor of $N_\rho$ from 
the  sum over $t_\rho$.  In this case only the I, V, and A channels
contribute; see \eq{dp-def}.  In the singlet case, we can use the fact 
that $M^2_{\eta',I}\approx m_0^2$  for large
$m_0$, and take the limit $m_0\to\infty$, resulting in one
less pole in the denominator of \eq{Disc} and an overall factor of $4/3$.  In the
vector and axial channels, we simply get a factor of $4$ from $N_\rho$.  This gives
the standard ratio $1/3$ between the strength of the singlet and vector or axial
hairpins, as seen in \cite{Aubin:2007mc}.  One may expect 
this standard ratio in any meson diagram,
such as \figref{tad-meson}(a),
that is unaffected by the complications from wrong spins or 
the naive-to-staggered translation.
The result is
\begin{eqnarray}
\cT_{x,{\rm \figrefeq{tad-quark-disca}}}^{(n)} &=&  -\frac{i}{f^2}
\Bigg\{
\frac{1}{3}\bigg[R^{[2,2]}_{X_I}\big(\{M^{(2)}_{X_I}\}
;\{\mu_I\}\big)\; \frac{\partial{\cal I}_{X,I}}{\partial
m^2_{X_I}}  
-\hspace{-3mm}\sum_{j \in
\{M^{(2)}_I\}}D^{[2,2]}_{j,X_I}\big(\{M^{(2)}_{X_I}\};\{\mu_I\}\big){\cal
I}_{j,I} \bigg]\nonumber \\
&&\hspace{5mm}
+a^2\delta'_{V}\bigg[R^{[3,2]}_{X_V}\big(\{M^{(3)}_{X_V}\}
;\{\mu_V\}\big)\; \frac{\partial{\cal I}_{X,V}}{\partial
m^2_{X_V}}   \nonumber
-\hspace{-3mm}\sum_{j \in
\{M^{(3)}_V\}}D^{[3,2]}_{j,X_V}\big(\{M^{(3)}_{X_V}\};\{\mu_V\}\big){\cal
I}_{j,V}
 \bigg]\nonumber\\  
&&\hspace{25mm} +\big(V\rightarrow A\big)\Bigg\}\ .
\eqn{T-disca}
\end{eqnarray}
Again, since wrong-spin contributions
to this diagram are absent, the result is independent of $n$.

\begin{figure}[t]
\begin{center}
\includegraphics[width=0.65\textwidth]{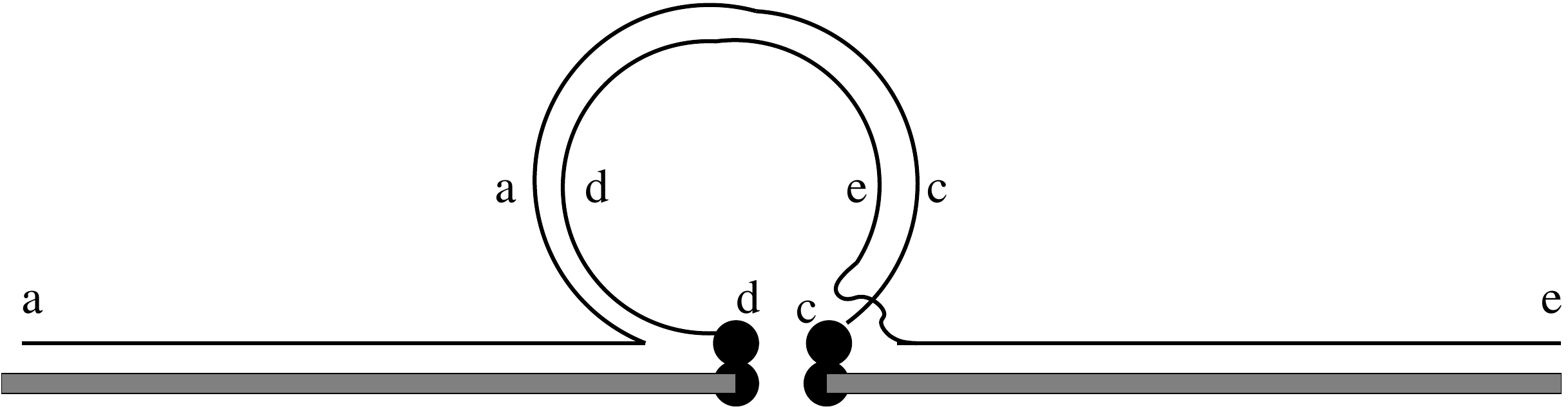}
\end{center}
\caption{Connected quark flow tadpole diagram.  This is a contribution to
\figref{tad-meson}(b). 
Taste-violations on the connected pion line allow for $a\not=c$ and $d\not=e$.
\label{fig:tad-quark-conn-x-x}}
\end{figure}

We now turn to contributions to the non-factorizable diagram, \figref{tad-meson}(b). 
The connected contribution
is shown in \figref{tad-quark-conn-x-x}. This is the first diagram in which
the quark flow connects the taste-$a$ quark in the external $P^{(b)}_{x,a}$
field with the light quark in the $P^{(\bar b)}$ field of the operator. According to the discussion above, this requires that we use the 
$c\leftrightarrow d$ versions of the operators in \eq{chiral-ops2}. The combination of taste matrices 
$\Gamma^{\rho,t_\rho}_{ad}\;  \Gamma^{\rho,t_\rho}_{ec}$
from the connected propagator \eq{PropConn}, and
$\Gamma^{\kappa,t_\kappa}_{ca}
\Gamma^{\kappa,t_\kappa}_{de}$ from the overall factor, \eq{factor2}, appears in
several diagrams.  It is therefore useful to define
\begin{equation}
\zeta(\kappa,\rho)
= \frac{1}{4N_\kappa N_\rho}\sum_{t_\kappa,t_\rho}{\rm tr}\left(\Gamma^{\kappa,t_\kappa}
\Gamma^{\rho,t_\rho}
\Gamma^{\kappa,t_\kappa}\;  \Gamma^{\rho,t_\rho}\right)\ .
\eqn{zeta}
\end{equation}
The factors of $N_\kappa$ and $N_\rho$ have been included in the denominator 
for later convenience.
Completeness of the
16 matrices $\Gamma^{\kappa,t_\kappa}$ implies that $\zeta$ satisfies the normalization condition,
\begin{equation}
\sum_{\kappa}N_\kappa\;\zeta(\kappa,\rho)
= 16\; \delta_{\rho,I}\ .
\eqn{zetanorm}
\end{equation}
Values of $\zeta(\kappa,\rho)$
are given in \tabref{zeta}.

\begin{table}[th]
\caption{Matrix of $\zeta(\kappa,\rho)$ values.}
\label{tab:zeta}
\begin{tabular}{|c|@{\hspace{8mm}}c@{\hspace{8mm}}c@{\hspace{8mm}}c@{\hspace{8mm}}c@{\hspace{8mm}}c|}
\hline
& $P$ & $A$ & $T$ & $V$ & $I$ \\
\hline
 $P$ & 1 & -1 & 1 & -1 & 1 \\
 $A$ & -1 & -1/2 & 0 & 1/2 & 1 \\
 $T$ & 1 &   0   & -1/3 & 0 & 1 \\
 $V$ & -1 & 1/2 & 0 & -1/2 & 1 \\
 $I$ & 1 & 1 & 1 & 1 & 1 \\
\hline
\end{tabular}
\end{table}

In terms of $\zeta(\kappa,\rho)$, the correct-spin and wrong-spin contributions
of \figref{tad-quark-conn-x-x} are then
\begin{eqnarray}
\cT_{x,{\rm \figrefeq{tad-quark-conn-x-x}}}^{(n)} &=&
\mp\frac{i}{16f^2}\sum_{\rho} \;N_\rho\; \cI_{X,\rho}\ ,
\eqn{T-conn-x-x} \\
\tilde\cT_{x,{\rm \figrefeq{tad-quark-conn-x-x}}}^{(n)}  &=&  
\mp\frac{i}{16f^2}\sum_{\kappa\not=I} \left(\frac{\beta_n^{(\kappa)}}{\beta_n}\;\sum_{\rho}N_\rho\;
\zeta(\kappa,\rho)\; \cI_{X,\rho}\right)\ ,
\eqn{Ttilde-conn-x-x}
\end{eqnarray}
where the upper sign is for $n=1,2,3$ and the lower sign is for $n=4,5$. The difference
in chiral structure of \eq{chiral-ops2} between $O^x_{4,5}$ and $O^x_{1,2,3}$ 
in \eq{chiral-ops2} matters for \figref{tad-meson}(b), because the two pion fields
come from different $\sigma$ or $\sigma^\dagger$ factors.
The correct-spin contribution $\cT_{x,{\rm Fig.5}}^{(n)}$ comes from the $\kappa=I$ term
in the sum, while the $\kappa\not=I$ terms give the wrong-spin piece $\tilde\cT_{x,{\rm Fig.5}}^{(n)}$.
The quantities $\beta_n^{(\kappa)}$ are listed in \tabref{beta}.

\begin{figure}[t]
\begin{center}
\includegraphics[width=0.65\textwidth]{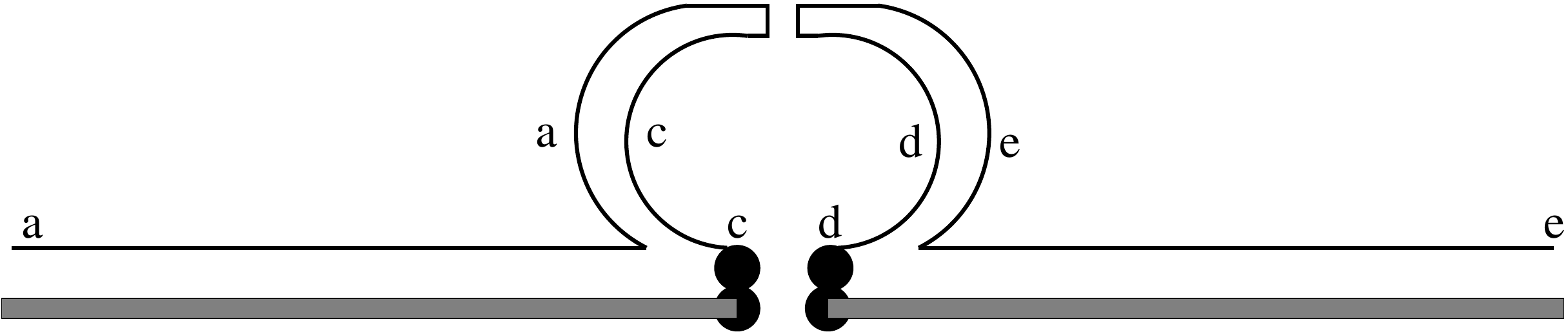}
\end{center}
\caption{Disconnected quark flow tadpole diagram.  This is a contribution to
\figref{tad-meson}(b). 
Taste-violating
hairpins allow $a\not=c$ and $d\not=e$.
\label{fig:tad-quark-discb}}
\end{figure}

The final tadpole diagram is the disconnected contribution to \figref{tad-meson}(b),
shown in \figref{tad-quark-discb}. In this case, the taste structure of the disconnected
propagator \eq{PropDisc} is\ \ $\Gamma^{\rho,t_\rho}_{ac}\Gamma^{\rho,t_\rho}_{ed}$, which, when
combined with the taste matrices in the overall factor, \eq{factor2}, 
gives ${\rm tr}^2(\Gamma^{\kappa,t_\kappa}\Gamma^{\rho,t_\rho})
=16\;\delta_{\kappa\rho}\,\delta_{t_\kappa t_\rho}$. The hairpin propagator is non-zero only
in the $I$, $V$, and $A$ channels. So we get a singlet contribution 
to the correct-spin operator 
($\kappa=\rho=I$) and taste-violating contributions for two of the wrong-spin operators
($\kappa=\rho=V,A$).  The result is 
\begin{eqnarray}
\cT_{x,{\rm \figrefeq{tad-quark-discb}}}^{(n)} &=&
\hspace{-1mm}\mp\frac{i}{3f^2}
\Bigg\{
R^{[2,2]}_{X_I}\big(\{M^{(2)}_{X_I}\}
;\{\mu_I\}\big)\; \frac{\partial{\cal I}_{X,I}}{\partial
m^2_{X_I}} 
-\hspace{-3mm}\sum_{j \in
\{M^{(2)}_I\}}D^{[2,2]}_{j,X_I}\big(\{M^{(2)}_{X_I}\};\{\mu_I\}\big){\cal
I}_{j,I} \Bigg\}\eqn{T-discb}\ , \\
\tilde\cT_{x,{\rm \figrefeq{tad-quark-discb}}}^{(n)}  &=&  
\hspace{-1mm}\mp\frac{i}{4\beta_nf^2}\Bigg\{
\beta_n^{(V)}\, a^2\delta'_{V}\bigg[R^{[3,2]}_{X_V}\big(\{M^{(3)}_{X_V}\}
;\{\mu_V\}\big)\; \frac{\partial{\cal I}_{X,V}}{\partial
m^2_{X_V}}   \nonumber \\
&&\hspace{40mm}-\hspace{-3mm}\sum_{j \in
\{M^{(3)}_V\}}D^{[3,2]}_{j,X_V}\big(\{M^{(3)}_{X_V}\};\{\mu_V\}\big){\cal
I}_{j,V}
 \bigg] \eqn{Ttilde-discb}\\
&&\hspace{3mm} 
+\beta_n^{(A)}\, a^2\delta'_{A}\bigg[R^{[3,2]}_{X_A}\big(\{M^{(3)}_{X_A}\}
;\{\mu_A\}\big)\; \frac{\partial{\cal I}_{X,A}}{\partial
m^2_{X_A}}  
-\hspace{-3mm}\sum_{j \in
\{M^{(3)}_A\}}D^{[3,2]}_{j,X_A}\big(\{M^{(3)}_{X_A}\};\{\mu_A\}\big){\cal
I}_{j,A}
 \bigg] 
\Bigg\}\ . \nonumber
\end{eqnarray}
Again the upper sign is for $n=1,2,3$ and the lower sign is for $n=4,5$; the reasoning
is the same as in \eqs{T-conn-x-x}{Ttilde-conn-x-x}.

\begin{figure}[t]
\begin{center}
\includegraphics[width=0.65\textwidth]{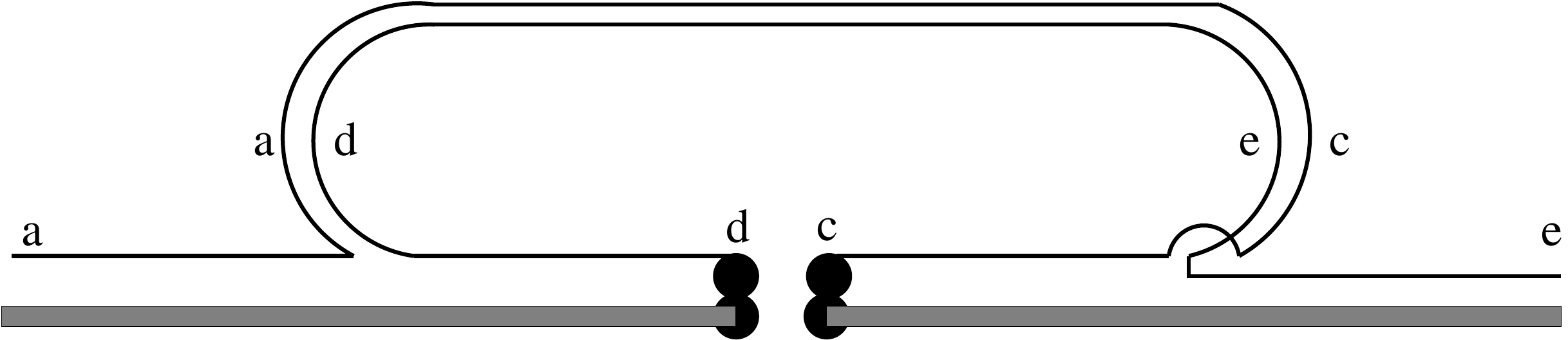}
\end{center}
\caption{Connected quark flow sunset graph corresponding
to  \figref{sunset-meson}.
Taste-violations on the connected pion line 
allow for $a\not=c$ and $d\not=e$.
\label{fig:sunset-quark-conn}}
\end{figure}

\begin{figure}[t]
\begin{center}
\includegraphics[width=0.65\textwidth]{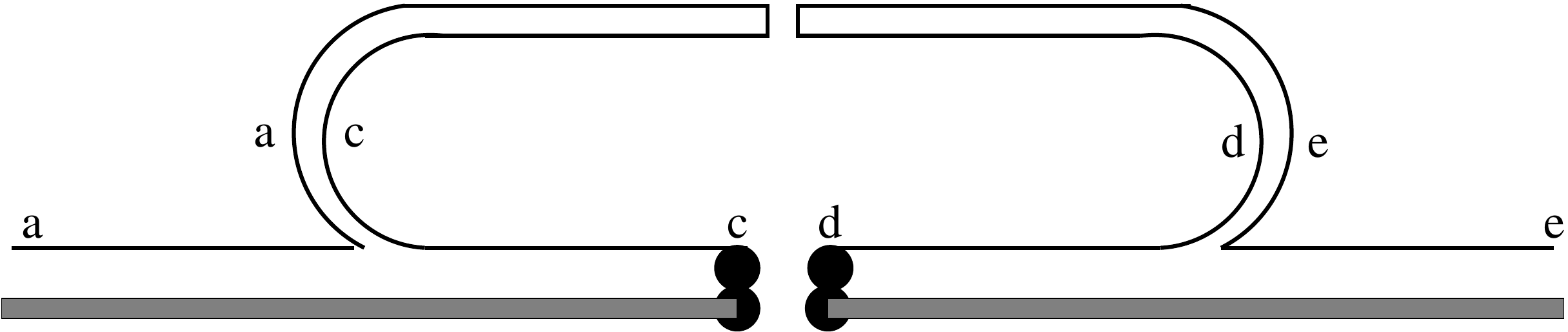}
\end{center}
\caption{Disconnected (hairpin) quark flow sunset graphs corresponding
to  \figref{sunset-meson}.
Taste-violating
hairpins allow for $a\not=c$ and $d\not=e$.
\label{fig:sunset-quark-disc}}
\end{figure}

The sunset diagram, \figref{sunset-meson}, is very similar to the tadpole contribution that
connects the incoming $B$ and outgoing $\bar B$, \figref{tad-meson}(b). Again, there
are two quark flows: \figref{sunset-quark-conn},  a connected graph similar to
\figref{tad-quark-conn-x-x}, and \figref{sunset-quark-disc}, a disconnected graph
similar to \figref{tad-quark-discb}. 
The taste structures of the sunset graphs are
identical to the corresponding tadpole graphs. 
As in \figref{tad-quark-conn-x-x},
the $c\leftrightarrow d$ version of \eq{chiral-ops2} is used in \figref{sunset-quark-conn}.  
The main difference between the sunset and tadpole graphs is the actual integral,
which here involves two heavy-light propagators and factors of $g_{B^*B\pi}$, and so
gives $g^2_{B^*B\pi}\cH^{\Delta^*}$ instead of $\cI$. In addition, the $\sigma$ and $\sigma^\dagger$
matrices in \eq{chiral-ops2} are all set to 1 in the sunset case, so there is no difference 
in overall
sign between the results for operators $1,2,3$ and those for operators $4,5$.
Otherwise, everything is the same as for the tadpole case.
 I find, for \figref{sunset-quark-conn},
\begin{eqnarray}
\cQ_{x,{\rm \figrefeq{sunset-quark-conn}}}^{(n)} &=&
-\frac{ig^2_{B^*B\pi}}{16f^2}\sum_{\rho} \;N_\rho\; \cH^{\Delta^*}_{X,\rho}\ ,
\eqn{Q-conn} \\
\tilde\cQ_{x,{\rm \figrefeq{sunset-quark-conn}}}^{(n)}  &=&  
-\frac{ig^2_{B^*B\pi}}{16f^2}\sum_{\kappa\not=I} \left( \frac{\beta_n^{\prime(\kappa)}}{\beta^\prime_n}\;\sum_{\rho}N_\rho\;
\zeta(\kappa,\rho)\; \cH^{\Delta^*}_{X,\rho}\right)\ .
\eqn{Qtilde-conn}
\end{eqnarray}

The disconnected sunset graph, \figref{sunset-quark-disc}, gives
\begin{eqnarray}
\cQ_{x,{\rm \figrefeq{sunset-quark-disc}}}^{(n)} &=&
\hspace{-1mm}\frac{-ig^2_{B^*B\pi}}{3f^2}
\Bigg\{
R^{[2,2]}_{X_I}\big(\{M^{(2)}_{X_I}\}
;\{\mu_I\}\big)\; \frac{\partial\cH^{\Delta^*}_{X,I}}{\partial
m^2_{X_I}} 
-\hspace{-3mm}\sum_{j \in
\{M^{(2)}_I\}}D^{[2,2]}_{j,X_I}\big(\{M^{(2)}_{X_I}\};\{\mu_I\}\big)\cH^{\Delta^*}_{j,I} 
\Bigg\}\eqn{Q-disc} \ , \\
\tilde\cQ_{x,{\rm \figrefeq{sunset-quark-disc}}}^{(n)}  &=&  
\hspace{-1mm}-\frac{ig^2_{B^*B\pi}}{4\beta'_n f^2}\Bigg\{
\beta^{\prime(V)}_n\, a^2\delta'_{V}\bigg[R^{[3,2]}_{X_V}\big(\{M^{(3)}_{X_V}\}
;\{\mu_V\}\big)\; \frac{\partial\cH^{\Delta^*}_{X,V}}{\partial
m^2_{X_V}}   \nonumber \\
&&\hspace{40mm}-\hspace{-3mm}\sum_{j \in
\{M^{(3)}_V\}}D^{[3,2]}_{j,X_V}\big(\{M^{(3)}_{X_V}\};\{\mu_V\}\big)\cH^{\Delta^*}_{j,V}
 \bigg] \eqn{Qtilde-disc}\\
&&\hspace{3mm} 
+\beta^{\prime(A)}_n\, a^2\delta'_{A}\bigg[R^{[3,2]}_{X_A}\big(\{M^{(3)}_{X_A}\}
;\{\mu_A\}\big)\; \frac{\partial\cH^{\Delta^*}_{X,A}}{\partial
m^2_{X_A}}\hspace{2mm}
-\hspace{-3mm}\sum_{j \in
\{M^{(3)}_A\}}\hspace{-2mm}
D^{[3,2]}_{j,X_A}\big(\{M^{(3)}_{X_A}\};\{\mu_A\}\big)\cH^{\Delta^*}_{j,A}
 \bigg] 
\Bigg\}\ . \nonumber
\end{eqnarray}

\subsection{Other Possible Taste-breaking Contributions}
\label{sec:taste-breaking}

Under renormalization, there are continuum-like 
mixings of the desired 4-quark lattice 
operators \cite{Evans:2009du}, but
there may in addition be perturbative
mixing with operators of incorrect spin and taste.
Further, there are discretization corrections to our identification
of the spin and taste of the operators, which
 can arise from variations in the heavy-quark field over a spatial cube,
as well as higher order terms
in the position-space spin-taste formalism \cite{TASTE-REPRESENTATION}.
However, incorrect spin-taste operators generated by any of these causes cannot
contribute to nonanalytic terms in the matrix element until NNLO.

In order to see this, consider the standard power-counting in (rooted) staggered chiral
perturbation theory ($rS\chi PT$):
\begin{equation}
    p^2 \sim m \sim a^2\  .
\end{equation}
The  $B^0-\bar B^0$ mixing four-fermion operators, when translated 
into the chiral effective theory, are of $\cO(1)$ in the 
aforementioned power-counting scheme.  Thus their LO (tree-level) 
contributions to the relevant matrix elements are of $\cO(1)$, 
and their NLO  (one-loop) contributions are of $\cO(p^2)$.

Perturbative mixing with wrong-taste operators
can occur at one-loop order in $\alpha_S$.  
In the chiral effective theory,
these wrong-taste operators would thus enter with 
coefficients of $\cO(\alpha_S/4\pi)$.  As was shown in Ref.~\cite{VandeWater:2005uq}, 
which considered the contribution of wrong-taste operators to neutral kaon mixing in \rschpt,
$\alpha_S/4\pi$ is numerically of the same size as the taste-breaking factor $a^2 \alpha_S^2$ on the $a\approx0.12$ fm MILC Asqtad ensembles \cite{RMP}.  
Thus the appropriate way to include the strong coupling constant in the \rhmschpt\ power-counting is:
\begin{equation}\eqn{power-counting}
    p^2 \sim m \sim a^2  \sim \alpha_S/4\pi\ .
\end{equation}
One-loop chiral diagrams involving the wrong-taste operators from perturbative mixing 
would therefore contribute to matrix elements 
only at NNLO, $\cO(\alpha_S/4\pi \,\, p^2)$, higher order than I am considering here.

Wrong-taste operators may also occur because 
of the $\cO(a)$ corrections
to the taste identification of the operators and the interpolating fields, 
\eqs{H-av2}{On-av}, 
coming either from the variation of
the heavy quark field over the 
hypercube (see \eq{H-av0}) or from $\cO(a)$ corrections to 
spin-taste identification of \eq{Kluberg}.
Since the matrix elements are taste conserving at tree level in \rhmschpt, any
such taste-violating effects must appear twice,
inducing $\cO(a^2)$ corrections.
These effects can then be absorbed into $a^2$-dependent 
analytic terms (see \eq{analytic} below).  Nontrivial
terms could appear at one loop in chiral perturbation theory, but the 
extra factor of $a^2$ implies that such terms are again effectively NNLO.

Consequently, only the LO taste breaking in the 4-quark operators,
together with taste-breaking terms in the 
LO pion  chiral Lagrangian, will modify the one-loop continuum chiral logarithms,
and these modification are what have been calculated above. 
Note however, that in more highly improved versions of staggered quarks,
taste violations (the $a^2$ terms in \eq{power-counting}) will be reduced
relative to the Asqtad case, but perturbative mixing may not be similarly reduced. In such
cases, it may be more reasonable to consider the $\cO(\alpha_S$) 
perturbative corrections to be of 
LO in the chiral expansion, and their one-loop chiral corrections
to be NLO, the same order as the 
corrections computed in \secref{Calculation}. I make some more comments
on this point in \secref{conclusions}.

\section{Final results}
\label{sec:results}

We can now combine results from different graphs to get the complete tadpole and
sunset contributions. For the correct-spin tadpole contribution, adding
\eqsfour{T-conn-x-s}{T-disca}{T-conn-x-x}{T-discb} gives
\begin{eqnarray}
 {\cal T}_{x}^{(1,2,3)} &=&
\frac{-i}{f_\pi^2}\Bigg\{\frac{1}{16}\sum_{\mathscr{S},\rho}
N_\rho\,{\cal I}_{x\mathscr{S},\rho}
+\frac{1}{16}\sum_{\rho}N_\rho\,{\cal I}_{X,\rho}
+\frac{2}{3}\bigg[R^{[2,2]}_{X_I}\big(\{M^{(2)}_{X_I}\}
;\{\mu_I\}\big)\; \frac{\partial{\cal I}_{X,I}}{\partial
m^2_{X_I}}  \nonumber
\\ && -\hspace{-3mm}\sum_{j \in
\{M^{(2)}_I\}}D^{[2,2]}_{j,X_I}\big(\{M^{(2)}_{X_I}\};\{\mu_I\}\big){\cal
I}_{j,I} \bigg]
+a^2\delta'_{V}\bigg[R^{[3,2]}_{X_V}\big(\{M^{(3)}_{X_V}\}
;\{\mu_V\}\big)\; \frac{\partial{\cal I}_{X,V}}{\partial
m^2_{X_V}}   \nonumber
\\ &&
-\hspace{-3mm}\sum_{j \in
\{M^{(3)}_V\}}D^{[3,2]}_{j,X_V}\big(\{M^{(3)}_{X_V}\};\{\mu_V\}\big){\cal
I}_{j,V}
 \bigg] +\big(V\rightarrow A\big)\Bigg\}, 
\eqn{T-correct-tot123}
\end{eqnarray}
\begin{eqnarray} 
{\cal T}^{(4,5)}_x &=&\frac{-i}{f_\pi^2}\Bigg\{\frac{1}{16}
\sum_{\mathscr{S},\rho}N_\rho\,{\cal I}_{x\mathscr{S},\rho}
-\frac{1}{16}\sum_{\rho}N_\rho\,  {\cal I}_{X,\rho}
+a^2\delta'_{V}\bigg[R^{[3,2]}_{X_V}\big(\{M^{(3)}_{X_V}\}
;\{\mu_V\}\big)\; \frac{\partial{\cal I}_{X,V}}{\partial
m^2_{X_V}}   \nonumber
\\ &&
-\hspace{-3mm}\sum_{j \in
\{M^{(3)}_V\}}D^{[3,2]}_{j,X_V}\big(\{M^{(3)}_{X_V}\};\{\mu_V\}\big){\cal
I}_{j,V}
 \bigg] +\big(V\rightarrow A\big)\Bigg\}. 
\eqn{T-correct-tot45}
\end{eqnarray}

The incorrect-spin tadpole contributions come from
\eqs{Ttilde-conn-x-x}{Ttilde-discb}. 
Adding them, and using the values 
of $\beta_n^{(\kappa)}$ in \tabref{beta} 
and of $\zeta(\kappa,\rho)$ in \tabref{zeta},
we have
\begin{eqnarray}
 \tilde {\cal T}_{x}^{(1)} &=&
\frac{-i}{f_\pi^2}\Bigg\{\frac{1}{16}
\bigg(-5{\cal I}_{X,P}-4{\cal I}_{X,A}+18{\cal I}_{X,T}-4{\cal I}_{X,V}-5{\cal I}_{X,I}\bigg) 
+\frac{2(\beta_2+\beta_3)}{\beta_1}\bigg(\hspace{-2mm}-\!{\cal I}_{X,V}+\!{\cal I}_{X,A}\nonumber\\
&&\hspace{5mm}+ a^2\delta'_{V}\Big[R^{[3,2]}_{X_V}\big(\{M^{(3)}_{X_V}\}
;\{\mu_V\}\big)\; \frac{\partial{\cal I}_{X,V}}{\partial
m^2_{X_V}}\hspace{3mm}
-\hspace{-3mm}\sum_{j \in
\{M^{(3)}_V\}}\hspace{-2mm}
D^{[3,2]}_{j,X_V}\big(\{M^{(3)}_{X_V}\};\{\mu_V\}\big){\cal
I}_{j,V}
 \Big]
\nonumber \\ 
&&\hspace{5mm}-a^2\delta'_{A}\Big[R^{[3,2]}_{X_A}\big(\{M^{(3)}_{X_A}\}
;\{\mu_A\}\big)\; \frac{\partial{\cal I}_{X,A}}{\partial
m^2_{X_A}} \hspace{3mm}
-\hspace{-3mm}\sum_{j \in
\{M^{(3)}_A\}}\hspace{-2mm}D^{[3,2]}_{j,X_A}\big(\{M^{(3)}_{X_A}\};\{\mu_A\}\big){\cal
I}_{j,A}
 \Big]
\bigg)\Bigg\}\ ,
\eqn{T1-wrong-tot}
\end{eqnarray}
\begin{eqnarray}
 \tilde {\cal T}_{x}^{(2)} &=&
\frac{-i}{f_\pi^2}\Bigg\{\frac{1}{16}
\bigg(-{\cal I}_{X,P}-4{\cal I}_{X,A}+10{\cal I}_{X,T}-4{\cal I}_{X,V}-{\cal I}_{X,I}\bigg) 
+\frac{\beta_3}{4\beta_2}\bigg(-{\cal I}_{X,P}+2{\cal I}_{X,T}-{\cal I}_{X,I}\bigg)\nonumber\\
&& \hspace{6mm} +\frac{\beta_1}{4\beta_2}\bigg(-{\cal I}_{X,V}+{\cal I}_{X,A}+
a^2\delta'_{V}\Big[R^{[3,2]}_{X_V}\big(\{M^{(3)}_{X_V}\}
;\{\mu_V\}\big)\;\frac{\partial{\cal I}_{X,V}}{\partial
m^2_{X_V}}  \nonumber \\
&&\hspace{15mm}-\hspace{-3mm}\sum_{j \in
\{M^{(3)}_V\}}\hspace{-2mm}D^{[3,2]}_{j,X_V}\big(\{M^{(3)}_{X_V}\};\{\mu_V\}\big){\cal
I}_{j,V}
 \Big]
-a^2\delta'_{A}\Big[R^{[3,2]}_{X_A}\big(\{M^{(3)}_{X_A}\}
;\{\mu_A\}\big)\; \frac{\partial{\cal I}_{X,A}}{\partial
m^2_{X_A}}  \nonumber \\
&&\hspace{15mm}-\hspace{-3mm}\sum_{j \in
\{M^{(3)}_A\}}\hspace{-2mm}
D^{[3,2]}_{j,X_A}\big(\{M^{(3)}_{X_A}\};\{\mu_A\}\big){\cal
I}_{j,A}
 \Big]
\bigg)\Bigg\}\ ,
\eqn{T2-wrong-tot}
\end{eqnarray}
\begin{eqnarray}
 \tilde {\cal T}_{x}^{(3)} &=&
\frac{-i}{f_\pi^2}\Bigg\{\frac{1}{16}
\bigg(-{\cal I}_{X,P}-4{\cal I}_{X,A}+10{\cal I}_{X,T}-4{\cal I}_{X,V}-{\cal I}_{X,I}\bigg) 
+\frac{\beta_2}{4\beta_3}\bigg(-{\cal I}_{X,P}+2{\cal I}_{X,T}-{\cal I}_{X,I}\bigg)\nonumber\\
&& \hspace{6mm} +\frac{\beta_1}{4\beta_3}\bigg(-{\cal I}_{X,V}+{\cal I}_{X,A}+
a^2\delta'_{V}\Big[R^{[3,2]}_{X_V}\big(\{M^{(3)}_{X_V}\}
;\{\mu_V\}\big)\;\frac{\partial{\cal I}_{X,V}}{\partial
m^2_{X_V}}  \nonumber \\
&&\hspace{15mm}-\hspace{-3mm}\sum_{j \in
\{M^{(3)}_V\}}\hspace{-2mm}D^{[3,2]}_{j,X_V}\big(\{M^{(3)}_{X_V}\};\{\mu_V\}\big){\cal
I}_{j,V}
 \Big]
-a^2\delta'_{A}\Big[R^{[3,2]}_{X_A}\big(\{M^{(3)}_{X_A}\}
;\{\mu_A\}\big)\; \frac{\partial{\cal I}_{X,A}}{\partial
m^2_{X_A}}  \nonumber \\
&&\hspace{15mm}-\hspace{-3mm}\sum_{j \in
\{M^{(3)}_A\}}\hspace{-2mm}
D^{[3,2]}_{j,X_A}\big(\{M^{(3)}_{X_A}\};\{\mu_A\}\big){\cal
I}_{j,A}
 \Big]
\bigg)\Bigg\}\ ,
\eqn{T3-wrong-tot}
\end{eqnarray}
\begin{eqnarray}
 \tilde {\cal T}_{x}^{(4)} &=&
\frac{i}{f_\pi^2}\Bigg\{\frac{1}{16}
\bigg(-{\cal I}_{X,P}+4{\cal I}_{X,A}-6{\cal I}_{X,T}+4{\cal I}_{X,V}-{\cal I}_{X,I}\bigg)
+\nonumber \\
&& \hspace{6mm} +\frac{\beta_5}{4\beta_4}\bigg({\cal I}_{X,P}-{\cal I}_{X,I} 
-2a^2\delta'_{V}\Big[R^{[3,2]}_{X_V}\big(\{M^{(3)}_{X_V}\}
;\{\mu_V\}\big)\;\frac{\partial{\cal I}_{X,V}}{\partial
m^2_{X_V}}  \nonumber \\
&&\hspace{15mm}-\hspace{-3mm}\sum_{j \in
\{M^{(3)}_V\}}\hspace{-2mm}D^{[3,2]}_{j,X_V}\big(\{M^{(3)}_{X_V}\};\{\mu_V\}\big){\cal I}_{j,V}
 \Big]
-2a^2\delta'_{A}\Big[R^{[3,2]}_{X_A}\big(\{M^{(3)}_{X_A}\}
;\{\mu_A\}\big)\; \frac{\partial{\cal I}_{X,A}}{\partial
m^2_{X_A}}  \nonumber \\
&&\hspace{15mm}-\hspace{-3mm}\sum_{j \in
\{M^{(3)}_A\}}\hspace{-2mm}
D^{[3,2]}_{j,X_A}\big(\{M^{(3)}_{X_A}\};\{\mu_A\}\big){\cal I}_{j,A}
 \Big]
\bigg)\Bigg\}\ ,
\eqn{T4-wrong-tot}
\end{eqnarray}
\begin{eqnarray}
 \tilde {\cal T}_{x}^{(5)} &=&
\frac{i}{f_\pi^2}\Bigg\{\frac{1}{16}
\bigg(-{\cal I}_{X,P}+4{\cal I}_{X,A}-6{\cal I}_{X,T}+4{\cal I}_{X,V}-{\cal I}_{X,I}\bigg)
+\nonumber \\
&& \hspace{6mm} +\frac{\beta_4}{4\beta_5}\bigg({\cal I}_{X,P}-{\cal I}_{X,I} 
-2a^2\delta'_{V}\Big[R^{[3,2]}_{X_V}\big(\{M^{(3)}_{X_V}\}
;\{\mu_V\}\big)\;\frac{\partial{\cal I}_{X,V}}{\partial
m^2_{X_V}}  \nonumber \\
&&\hspace{15mm}-\hspace{-3mm}\sum_{j \in
\{M^{(3)}_V\}}\hspace{-2mm}D^{[3,2]}_{j,X_V}\big(\{M^{(3)}_{X_V}\};\{\mu_V\}\big){\cal I}_{j,V}
 \Big]
-2a^2\delta'_{A}\Big[R^{[3,2]}_{X_A}\big(\{M^{(3)}_{X_A}\}
;\{\mu_A\}\big)\; \frac{\partial{\cal I}_{X,A}}{\partial
m^2_{X_A}}  \nonumber \\
&&\hspace{15mm}-\hspace{-3mm}\sum_{j \in
\{M^{(3)}_A\}}\hspace{-2mm}
D^{[3,2]}_{j,X_A}\big(\{M^{(3)}_{X_A}\};\{\mu_A\}\big){\cal I}_{j,A}
 \Big]
\bigg)\Bigg\}\ .
\eqn{T5-wrong-tot}
\end{eqnarray}
In the continuum limit,
when all taste propagators become degenerate and $a^2\delta'_{V,A}\to0$,  the wrong-spin
terms clearly vanish.

For the correct-spin sunset diagrams, adding \eqs{Q-conn}{Q-disc} gives
\begin{eqnarray} 
{\cal Q}^{(n)}_{x} &=&
\frac{-ig_{B^*B\pi}^2}{f_\pi^2}\Bigg\{\frac{1}{16}\sum_{ \rho}
N_\rho\,{\cal H}^{\Delta^*}_{X,\rho}
+\frac{1}{3}\bigg[R^{[2,2]}_{X_I}\big(\{M^{(2)}_{X_I}\}
;\{\mu_I\}\big)\; \frac{\partial{{\cal
H}^{\Delta^*}_{X,I}}}{\partial m^2_{X_I}}  \nonumber
\\ && -\hspace{-3mm}\sum_{j \in
\{M^{(2)}_I\}}D^{[2,2]}_{j,X_I}\big(\{M^{(2)}_{X_I}\};\{\mu_I\}\big){\cal
H}^{\Delta^*}_{j,I} \bigg] \Bigg\}\ . 
\eqn{Q-correct-tot}
\end{eqnarray}

The incorrect-spin sunset contributions come from
\eqs{Qtilde-conn}{Qtilde-disc}. 
Again using the values of $\zeta{(\kappa,\rho)}$
in \tabref{zeta} and of $\beta_n^{(\kappa)}$ in \tabref{beta}, we have
\begin{eqnarray} 
\tilde{\cal Q}^{(1)}_{x} &=&
\frac{-ig_{B^*B\pi}^2}{f_\pi^2}\Bigg\{
\frac{1}{16}
\bigg(-5{\cal H}^{\Delta^*}_{X,P}-4{\cal H}^{\Delta^*}_{X,A}+18{\cal H}^{\Delta^*}_{X,T}-4{\cal H}^{\Delta^*}_{X,V}-5{\cal H}^{\Delta^*}_{X,I}\bigg) \nonumber \\
&&\hspace{12mm}+\frac{2(\beta'_2+\beta'_3)}{\beta_1}\Big(-{\cal H}^{\Delta^*}_{X,V}+{\cal H}^{\Delta^*}_{X,A}+ \nonumber\\
&&\hspace{-7mm}+ a^2\delta'_{V}\Big[R^{[3,2]}_{X_V}\big(\{M^{(3)}_{X_V}\}
;\{\mu_V\}\big)\; \frac{\partial{\cal H}^{\Delta^*}_{X,V}}{\partial
m^2_{X_V}}   
-\hspace{-3mm}\sum_{j \in
\{M^{(3)}_V\}}\hspace{-3mm}D^{[3,2]}_{j,X_V}\big(\{M^{(3)}_{X_V}\};\{\mu_V\}\big){\cal
H}^{\Delta^*}_{j,V}
 \Big]
\nonumber \\ 
&&\hspace{-7mm}-a^2\delta'_{A}\Big[R^{[3,2]}_{X_A}\big(\{M^{(3)}_{X_A}\}
;\{\mu_A\}\big)\; \frac{\partial{\cal H}^{\Delta^*}_{X,A}}{\partial
m^2_{X_A}} 
-\hspace{-3mm}\sum_{j \in
\{M^{(3)}_A\}}\hspace{-4mm}D^{[3,2]}_{j,X_A}\big(\{M^{(3)}_{X_A}\};\{\mu_A\}\big){\cal H}^{\Delta^*}
_{j,A}
 \Big]
\bigg)\Bigg\}\ ,
\eqn{Q1-wrong-tot}
\end{eqnarray}
\begin{eqnarray}
 \tilde {\cal Q}_{x}^{(2)} &=&
\frac{-ig_{B^*B\pi}^2}{f_\pi^2}\Bigg\{\frac{1}{16}
\bigg(-{\cal H}^{\Delta^*}_{X,P}-4{\cal H}^{\Delta^*}_{X,A}+10{\cal H}^{\Delta^*}_{X,T}-4{\cal H}^{\Delta^*}_{X,V}-{\cal H}^{\Delta^*}_{X,I}\bigg) \nonumber \\
&&\hspace{6mm}+\frac{\beta'_3}{4\beta'_2}\bigg(-{\cal H}^{\Delta^*}_{X,P}+2{\cal H}^{\Delta^*}_{X,T}-{\cal H}^{\Delta^*}_{X,I}\bigg)
+\frac{\beta_1}{4\beta'_2}\bigg(-{\cal H}^{\Delta^*}_{X,V}+{\cal H}^{\Delta^*}_{X,A}+ \nonumber\\
&&\hspace{-7mm}+a^2\delta'_{V}\Big[R^{[3,2]}_{X_V}\big(\{M^{(3)}_{X_V}\}
;\{\mu_V\}\big)\;\frac{\partial{\cal H}^{\Delta^*}_{X,V}}{\partial
m^2_{X_V}}  
-\hspace{-3mm}\sum_{j \in
\{M^{(3)}_V\}}\hspace{-2mm}D^{[3,2]}_{j,X_V}\big(\{M^{(3)}_{X_V}\};\{\mu_V\}\big){\cal H}^{\Delta^*}_{j,V}
 \Big]\nonumber \\
&&\hspace{-7mm}-a^2\delta'_{A}\Big[R^{[3,2]}_{X_A}\big(\{M^{(3)}_{X_A}\}
;\{\mu_A\}\big)\; \frac{\partial{\cal H}^{\Delta^*}_{X,A}}{\partial
m^2_{X_A}}  
-\hspace{-3mm}\sum_{j \in
\{M^{(3)}_A\}}\hspace{-2mm}
D^{[3,2]}_{j,X_A}\big(\{M^{(3)}_{X_A}\};\{\mu_A\}\big){\cal H}^{\Delta^*}_{j,A}
 \Big]
\bigg)\Bigg\}\ ,
\eqn{Q2-wrong-tot}
\end{eqnarray}
\begin{eqnarray}
 \tilde {\cal Q}_{x}^{(3)} &=&
\frac{-ig_{B^*B\pi}^2}{f_\pi^2}\Bigg\{\frac{1}{16}
\bigg(-{\cal H}^{\Delta^*}_{X,P}-4{\cal H}^{\Delta^*}_{X,A}+10{\cal H}^{\Delta^*}_{X,T}-4{\cal H}^{\Delta^*}_{X,V}-{\cal H}^{\Delta^*}_{X,I}\bigg) \nonumber \\
&&\hspace{6mm}+\frac{\beta'_2}{4\beta'_3}\bigg(-{\cal H}^{\Delta^*}_{X,P}+2{\cal H}^{\Delta^*}_{X,T}-{\cal H}^{\Delta^*}_{X,I}\bigg)
+\frac{\beta_1}{4\beta'_3}\bigg(-{\cal H}^{\Delta^*}_{X,V}+{\cal H}^{\Delta^*}_{X,A}+\nonumber \\
&&\hspace{-7mm}+a^2\delta'_{V}\Big[R^{[3,2]}_{X_V}\big(\{M^{(3)}_{X_V}\}
;\{\mu_V\}\big)\;\frac{\partial{\cal H}^{\Delta^*}_{X,V}}{\partial
m^2_{X_V}}  
-\hspace{-3mm}\sum_{j \in
\{M^{(3)}_V\}}\hspace{-2mm}D^{[3,2]}_{j,X_V}\big(\{M^{(3)}_{X_V}\};\{\mu_V\}\big){\cal H}^{\Delta^*}_{j,V}
 \Big] \nonumber \\
&&\hspace{-7mm} -a^2\delta'_{A}\Big[R^{[3,2]}_{X_A}\big(\{M^{(3)}_{X_A}\}
;\{\mu_A\}\big)\; \frac{\partial{\cal H}^{\Delta^*}_{X,A}}{\partial
m^2_{X_A}}  
-\hspace{-3mm}\sum_{j \in
\{M^{(3)}_A\}}\hspace{-2mm}
D^{[3,2]}_{j,X_A}\big(\{M^{(3)}_{X_A}\};\{\mu_A\}\big){\cal H}^{\Delta^*}_{j,A}
 \Big]
\bigg)\Bigg\}\ ,
\eqn{Q3-wrong-tot}
\end{eqnarray}
\begin{eqnarray}
 \tilde {\cal Q}_{x}^{(4)} &=&
\frac{-ig_{B^*B\pi}^2}{f_\pi^2}\Bigg\{\frac{1}{16}
\bigg(-{\cal H}^{\Delta^*}_{X,P}+4{\cal H}^{\Delta^*}_{X,A}-6{\cal H}^{\Delta^*}_{X,T}+4{\cal H}^{\Delta^*}_{X,V}-{\cal H}^{\Delta^*}_{X,I}\bigg)+ \nonumber\\
&& \hspace{15mm} +\frac{\beta'_5}{4\beta'_4}\bigg({\cal H}^{\Delta^*}_{X,P}-{\cal H}^{\Delta^*}_{X,I} 
-2a^2\delta'_{V}\Big[R^{[3,2]}_{X_V}\big(\{M^{(3)}_{X_V}\}
;\{\mu_V\}\big)\;\frac{\partial{\cal H}^{\Delta^*}_{X,V}}{\partial
m^2_{X_V}}  \nonumber \\
&&\hspace{5mm}-\hspace{-3mm}\sum_{j \in
\{M^{(3)}_V\}}\hspace{-2mm}D^{[3,2]}_{j,X_V}\big(\{M^{(3)}_{X_V}\};\{\mu_V\}\big){\cal H}^{\Delta^*}_{j,V}
 \Big]
-2a^2\delta'_{A}\Big[R^{[3,2]}_{X_A}\big(\{M^{(3)}_{X_A}\}
;\{\mu_A\}\big)\; \frac{\partial{\cal H}^{\Delta^*}_{X,A}}{\partial
m^2_{X_A}}  \nonumber \\
&&\hspace{10mm}-\hspace{-3mm}\sum_{j \in
\{M^{(3)}_A\}}\hspace{-2mm}
D^{[3,2]}_{j,X_A}\big(\{M^{(3)}_{X_A}\};\{\mu_A\}\big){\cal H}^{\Delta^*}_{j,A}
 \Big]
\bigg)\Bigg\}\ ,
\eqn{Q4-wrong-tot}
\end{eqnarray}
\begin{eqnarray}
 \tilde {\cal Q}_{x}^{(5)} &=&
\frac{-ig_{B^*B\pi}^2}{f_\pi^2}\Bigg\{\frac{1}{16}
\bigg(-{\cal H}^{\Delta^*}_{X,P}+4{\cal H}^{\Delta^*}_{X,A}-6{\cal H}^{\Delta^*}_{X,T}+4{\cal H}^{\Delta^*}_{X,V}-{\cal H}^{\Delta^*}_{X,I}\bigg)+ \nonumber\\
&& \hspace{15mm} +\frac{\beta'_4}{4\beta'_5}\bigg({\cal H}^{\Delta^*}_{X,P}-{\cal H}^{\Delta^*}_{X,I} 
-2a^2\delta'_{V}\Big[R^{[3,2]}_{X_V}\big(\{M^{(3)}_{X_V}\}
;\{\mu_V\}\big)\;\frac{\partial{\cal H}^{\Delta^*}_{X,V}}{\partial
m^2_{X_V}}  \nonumber \\
&&\hspace{5mm}-\hspace{-3mm}\sum_{j \in
\{M^{(3)}_V\}}\hspace{-2mm}D^{[3,2]}_{j,X_V}\big(\{M^{(3)}_{X_V}\};\{\mu_V\}\big){\cal H}^{\Delta^*}_{j,V}
 \Big]
-2a^2\delta'_{A}\Big[R^{[3,2]}_{X_A}\big(\{M^{(3)}_{X_A}\}
;\{\mu_A\}\big)\; \frac{\partial{\cal H}^{\Delta^*}_{X,A}}{\partial
m^2_{X_A}}  \nonumber \\
&&\hspace{10mm}-\hspace{-3mm}\sum_{j \in
\{M^{(3)}_A\}}\hspace{-2mm}
D^{[3,2]}_{j,X_A}\big(\{M^{(3)}_{X_A}\};\{\mu_A\}\big){\cal H}^{\Delta^*}_{j,A}
 \Big]
\bigg)\Bigg\}\ .
\eqn{Q5-wrong-tot}
\end{eqnarray}

In comparing \eqsfour{wave-function}{T-correct-tot123}{T-correct-tot45}{Q-correct-tot}
to the continuum results of 
Ref.~\cite{Detmold:2006gh}, one needs to be aware of the many differences
in notation. In particular, the particles called
$X$ and $\pi$ in Ref.~\cite{Detmold:2006gh} are called $\eta$ and $X$, respectively, here. 
Taking the notational differences into account, it is straightforward
to check that the continuum limits of the above equations reproduce 
the results in \rcite{Detmold:2006gh}.

The analytic terms in \eqs{O1tot}{O2-5tot} are of the form
\begin{equation} \eqn{analytic}
\textrm{analytic terms} = L^{(n)}_v m_x + L^{(n)}_s (2m_l +m_h) + L^{(n)}_a a^2\,, 
\end{equation}
where $L^{(n)}_v$ and $L^{(n)}_s$ are continuum low-energy constants, and $L^{(n)}_a$ 
are lattice low-energy constants, summarizing the
effects of taste-violating analytic chiral operators at NLO. 
As indicated, these constants
depend on $n$, \ie on the operator whose matrix element is being
calculated.  It is straightforward to see that these are in fact the analytic
terms that may appear by considering the effects of
adding mass or taste-violating spurions to the chiral operators in 
\eq{chiral-ops2}.
The parameters $L^{(n)}_v$, $L^{(n)}_s$, and $L^{(n)}_a$, 
together with the parameters
$\hat\beta_n$ and $\hat{\beta'}_n$
in \eqs{O1tot}{O2-5tot}
are to be determined from the fits to lattice data. 

\section{Conclusions and Discussion}
\label{sec:conclusions}

I have calculated neutral $B$ mixing to one-loop in staggered chiral perturbation theory for the complete set of Standard Model and beyond-the-Standard-Model
operators, \eq{SUSY-basis}.
My results are given by \eqs{O1tot}{O2-5tot}, with expressions
for the various terms listed in \eq{wave-function} and
\eqsthru{T-correct-tot123}{analytic}.

The construction of the operators on the lattice
as local products of local heavy-light bilinears, coupled with the use of
staggered light quarks, results in the appearance of
``wrong spin/taste operators'' that
are $\cO(1)$ in the lattice spacing.  
Their contributions to the matrix elements considered here
are suppressed to NLO because violation of
taste symmetry is required. At that order, they 
induce mixing of the operators, summarized in the quantities
$\tilde T^{(n)}$, \eqsthru{T1-wrong-tot}{T5-wrong-tot}, and $\tilde Q^{(n)}$,
\eqsthru{Q1-wrong-tot}{Q5-wrong-tot}.   However,
as long as all five $\cO_n$ are analyzed simultaneously,
there are no new low-energy constants induced by these effects: the 
constants $\beta_n$ and $\beta'_n$ are all already present in the continuum.

Effective operator mixings may come from three additional sources,
 weak-coupling perturbative corrections, corrections to
the position-space spin-taste construction, and corrections to the taste identification
of the operators and interpolating fields. 
I argue that any nonanalytic terms that arise from such mixings
are effectively NNLO, higher order than what has been considered here.

Because the light staggered quark  is converted to a naive quark in the
lattice representatives of the operators, the relationship between the
staggered and naive quark fields plays an important role in my analysis.
In particular, a naive quark is equivalent to four copies of staggered
quarks, and the resulting ``copy symmetry'' can be used to simplify the
calculations.  An interesting resulting feature is that the taste structure
of the operators is not the same for all diagrams, but depends on the quark flow.

Lattice computations in Refs.~\cite{Gamiz:2009ku,Bazavov:2012zs} 
focused on the calculation of the quantity $\xi 
\equiv (f_{B_s}\sqrt{\hat B_{B_s}})/( f_{B_d}\sqrt{\hat B_{B_d}})$, which comes from the matrix element of operator $\cO_1$.
The chiral effects of the wrong spins were not known at the time
of the HPQCD calculation \cite{Gamiz:2009ku} and were therefore omitted from the analysis and error estimate.
In the Fermilab/MILC calculation \cite{Bazavov:2012zs}, the complete \rhmschpt\ expressions
were available, but the
matrix elements of operators other than $\cO_1$ were not calculated, preventing a direct
inclusion of the wrong-spin effects.  However, it was possible to estimate the error
of omitting these effects by using
a small subset of new data to investigate the other matrix elements.
The result, $\xi = 1.268(63)$  included a 3.2\% error from this effect, which was
the second largest source of error.
In the ongoing second-generation Fermilab/MILC 
project \cite{Freeland:2012kz}, matrix elements
of all five operators $\cO_n$ are being computed, which means that the complete
\rhmschpt\ expressions  can be used in the analysis, and there will be no ``wrong-spin
error.'' Of course, a chiral/continuum extrapolation error will remain.

For future lattice computations of mixing with 
``highly improved staggered quark'' (HISQ) ensembles \cite{HISQ}, taste
violations are sufficiently reduced that the power counting used
here, \eq{power-counting},
may no longer be appropriate. Depending on the range of lattice
spacings studied and the statistical errors in the data, 
taste violations may in fact be so small that
continuum heavy meson \chpt\ might prove adequate for describing the data.  More likely,
one will want to use the \rhmschpt\ forms calculated here, but it may 
be necessary in addition to include 
the NLO chiral effects of the
operators that enter through weak-coupling perturbative mixing, since such effects
may no longer be much smaller than taste-violating NLO effects.
Including such effects in NLO \rhmschpt\ will however be straightforward, 
since the chiral logarithms for the
complete set of operators, \eq{SUSY-basis}, have already been calculated
above. 

\section{Acknowledgments}

I am indebted to J.\ Laiho, R.S.\ Van de Water, and C.\ Bouchard for
help with various aspects of the calculation, and thank A.\ Kronfeld and all my other
colleagues in the
MILC and Fermilab/Lattice Collaborations for discussions.
This work has been partially supported by the Department of Energy, under grant number DE-FG02-91ER40628. I also thank the Galileo
Galilei Institute for Theoretical Physics for the hospitality
and the INFN for partial support while this work was in progress.

\appendix
\section{\label{sec:Bparams} B Parameters}

It is sometimes convenient to express the mixing matrix elements in terms
of B (or ``bag'') parameters.  A fairly common set of definitions is given
for example in Ref.~\cite{Detmold:2006gh}:

\begin{eqnarray}
\eqn{B1}
\langle \overline{B}_x^0|{O}_1^x|B_x^0 \rangle &=& \frac{8}{3} M^2_{B_x} f^2_{B_x}B_{B_x}^{(1)}\;,\\
\langle \overline{B}_x^0|{O}_n^x|B_x^0 \rangle &=&
\eta_n R^2 M^2_{B_x} f^2_{B_x} B_{B_x}^{(n)}\quad\textrm{for\ }n=2,3,4,5\;,
\eqn{B2-5}
\end{eqnarray}
where $f_{B_x}$ is the decay constant of the $B_x$ meson, 
$M_{B_x}$ is its mass, $R\equiv M_{B_x}/(m_b+m_x)$, and
$\eta_2=-5/3$, $\eta_3=1/3$, $\eta_4=2$, and $\eta_5=2/3$. Relativistic normalization
of the states is assumed in these expressions.

The expression for the decay constant in \rhmschpt, including heavy-meson
hyperfine and flavor splittings, is given in
Eq.~(6.20) of Ref.~\cite{Bazavov:2011aa}. To convert it to the current
notation, we just must replace the chiral logarithm functions 
$J$ and $\ell$ with $\cH$ and $\cI$ using \eqs{H-J}{I-ell}. Using
quantities defined above in \eqsthree{wave-function}{T-conn-x-s}{T-disca},
we may write the result as
\begin{equation}\eqn{Phi}
f_{B_x}\sqrt{M_{B_x}} = \Phi_0 \left[ 1 + \frac{1}{2}\left( \cW_{x\bar b}
+\cT_{x,{\rm \figrefeq{tad-quark-conn-x-s}}}^{(n)}
+ \cT_{x,{\rm \figrefeq{tad-quark-disca}}}^{(n)} \right)\right]\;.
\end{equation}
It is not surprising that $\cW_{x\bar b}$, $\cT_{x,{\rm \figrefeq{tad-quark-conn-x-s}}}^{(n)}$, and $ \cT_{x,{\rm \figrefeq{tad-quark-disca}}}^{(n)}$ appear,
because the wave function and the tadpole contributions of \figrefs{tad-quark-conn-x-s}{tad-quark-disca} are factorizable: they affect only one meson and
one bilinear of the 4-quark operators, so they are exactly the contributions
that appear in the decay constant.

As always, 
heavy-light chiral perturbation theory is expressed in terms
of the nonrelativistically normalized states of heavy quark effective theory.
Thus the low energy constant, $\Phi_0$, that describes the decay constant
at tree level in chiral perturbation theory includes a factor of $\sqrt{M_{B_x}}$.  
Similarly, with the relativistic normalization used \eqs{B1}{B2-5}, 
one factor of $M_{B_x}$ needs to be included in our expressions
for these matrix elements in terms of the
parameters $\beta_n$ and $\beta'_n$. Taking these normalization factors into account, and
using \eqsthree{O1tot}{O2-5tot}{Phi}, we have
\begin{eqnarray}
\eqn{B1res}
B_{B_x}^{(1)} &=&
\frac{\beta_1}{(8/3)\Phi^2_0}\left(1+
{\cal S}_x + \tilde{\cal T}^{(1)}_x 
+ {\cal Q}^{(1)}_x+ \tilde{\cal Q}^{(1)}_x\right)
+  \textrm{analytic terms} \\
B_{B_x}^{(n)} &=&
\frac{\beta_n}{\eta_nR^2\Phi^2_0}\left(1
\pm{\cal S}_x + \tilde{\cal T}^{(n)}_x\right)
+ \frac{\beta'_n}{\eta_nR^2\Phi^2_0}
\left({\cal
Q}^{(n)}_x+ \tilde{\cal Q}^{(n)}_x\right)+ \textrm{analytic terms}\;,
\eqn{B2-5res}
\end{eqnarray}
where $n=2,\dots,5$ in the second equation, and the upper (plus) sign is for
$n=2,3$, while the lower (minus) sign is for $n=4,5$. 
Here the wave function and tadpole contributions of 
\figrefs{tad-quark-conn-x-s}{tad-quark-disca} have canceled, and $\cS_x$ is a new correct-spin
tadpole term that comes only from the nonfactorizable tadpole diagrams,
\figrefs{tad-quark-conn-x-x}{tad-quark-discb}. 
Combining
\eqs{T-conn-x-x}{T-discb} gives  
\begin{eqnarray}
 {\cal S}_{x} &=& 
\frac{-i}{f_\pi^2}\Bigg\{
\frac{1}{16}\sum_{\rho}N_\rho\,{\cal I}_{X,\rho}+ \nonumber \\
&&\hspace{6mm}+\frac{1}{3}\bigg[R^{[2,2]}_{X_I}\big(\{M^{(2)}_{X_I}\}
;\{\mu_I\}\big)\; \frac{\partial{\cal I}_{X,I}}{\partial
m^2_{X_I}}\hspace{1mm}-\hspace{-3mm}
\sum_{j \in
\{M^{(2)}_I\}}\hspace{-3mm}D^{[2,2]}_{j,X_I}\big(\{M^{(2)}_{X_I}\};\{\mu_I\}\big){\cal
I}_{j,I} \bigg] \Bigg\}\;. 
\eqn{S}
\end{eqnarray}
The other chiral logarithm functions in \eqs{B1res}{B2-5res} 
are given in \eqsthru{T1-wrong-tot}{Q5-wrong-tot}.
The analytic terms in \eqs{B1res}{B2-5res} have the same form as in \eq{analytic} (with,
of course, redefined low energy constants).

\end{document}